\documentclass[aps,prd,amsmath,amsfonts,amssymb,twocolumn,nofootinbib,showkeys]{revtex4}

\usepackage{amssymb}
\usepackage{graphicx,psfrag,color}
\usepackage{dcolumn}
\usepackage{bm}
\usepackage{mathbbol}
\usepackage{slashed}
\usepackage[utf8]{inputenc}
\DeclareUnicodeCharacter{2009}{\,} 

\begin{document}
\flushbottom

\title{Asymmetric particle-antiparticle Dirac equation: second quantization}

\author{Gustavo Rigolin}
\email{rigolin@ufscar.br}
\affiliation{Departamento de F\'isica, Universidade Federal de
S\~ao Carlos, 13565-905, S\~ao Carlos, S\~ao Paulo, Brazil}

\date{\today}

\begin{abstract}
We build the fully relativistic quantum field theory related to the asymmetric Dirac fields first presented in a prequel to this work. 
These fields are solutions of the asymmetric Dirac equation, 
a Lorentz covariant Dirac-like equation whose positive and ``negative'' frequency plane wave solutions' dispersion relations are no longer degenerate. At the second quantization 
level, we show that this implies that particles and antiparticles sharing the same wave number have different energies and momenta. In spite of that, we prove that by
properly fixing the values of the relativistic invariants that define the 
asymmetric Dirac free field Lagrangian density, we can build a consistent, fully relativistic, and renormalizable quantum electrodynamics (QED) that is empirically 
equivalent to the standard QED. We discuss the reasons and implications of this non-trivial
equivalence, exploring qualitatively other scenarios in which the asymmetric Dirac fields
may lead to beyond the standard model predictions. 
We give a complete account of 
how the asymmetric Dirac fields and the corresponding annihilation and creation operators transform under improper Lorentz transformations (parity and time reversal operations) and under the charge conjugation operation. We also prove that the present theory  
respects the CPT theorem. 
\end{abstract}



\maketitle

\section{Introduction} 

Strictly speaking, the non-relativistic Schr\"odinger wave function $\psi(x)$ does not 
transform as a scalar if we demand the covariance of
the Schr\"odinger equation after a Galilean boost. The covariance of the Schr\"odinger
equation \cite{sch26} is only achieved if the wave function transforms as follows 
after a Galilean boost \cite{bal98},
\begin{equation}
\psi(x) = e^{\frac{i}{\hbar}\theta(x')} \psi'(x'), 
\label{psiStransform}
\end{equation}
where $\psi(x)$ and $\psi'(x')$ are the wave functions describing the same physical system
in two inertial reference frames $S$ and $S'$, with the latter moving away from the former
with velocity $\mathbf{v}$, and 
\begin{equation}
\theta(x') = \frac{mv^2}{2}t' + m \mathbf{v\cdot\mathbf r'}.
\label{thetaS}
\end{equation}

This should be compared with the Klein-Gordon wave function \cite{gre00},
a strict scalar whose transformation law after a Lorentz boost is 
\begin{equation}
\phi(x) = \phi'(x'), 
\label{phiStransform}
\end{equation}
where $x=(ct,\mathbf{r})$ and $x'=(ct',\mathbf{r'})$. 

In Eqs.~(\ref{psiStransform})-(\ref{phiStransform}), 
$t(t')$ and $\mathbf{r}(\mathbf{r'})$ are the time
and position of the particle in $S(S')$, $v^2=|\mathbf{v}|^2=\mathbf{v\cdot v}$,
$m$ is the mass of the particle, $\hbar$ is Planck's constant divided by $2\pi$,
and $c$ is the speed of light. 

In Ref.~\cite{rig22} we obtained the Lorentz covariant wave equation associated with
the most natural relativistic extension of the transformation law (\ref{psiStransform}).
It turned out that this wave equation
had first and second order derivatives. 
In Ref.~\cite{rig23} we adapted Dirac's approach 
\cite{gre00,dir67,man86,gre95,sch05,bog80} to obtain a first order differential equation 
whose ``square'' recovered the Lorentz covariant wave equation 
first derived in Ref.~\cite{rig22}. The wave equation presented in Ref.~\cite{rig23} is,
similarly to Dirac's, a spinorial homogeneous first order differential equation.
However, particles and antiparticles
have different energy and momentum for a given wave number. As such, we called it
\textit{asymmetric Dirac equation}.

An extensive investigation of the first quantized properties of the asymmetric Dirac
equation was made in Ref.~\cite{rig23}, including the derivation of the  
transformation law of its wave function after proper and improper Lorentz transformations 
and the proof of its covariance under those transformations.

In this work, a sequel of Ref.~\cite{rig23}, 
our main goal is to second quantize the asymmetric Dirac equation and
show that a consistent quantum field theory can be built out of it. In particular,
we show that by properly fixing the free parameters of the asymmetric Dirac equation
and minimally coupling it with the electromagnetic field, a renormalizable 
quantum electrodynamics (QED) theory giving the same predictions of the standard 
QED can be built. 

In order to be as self-contained as possible, in Sec.~\ref{asd} we review the main properties of the asymmetric Dirac equation and
its Lagrangian formulation \cite{rig23}, which also helps in 
setting the notation that will be used in the rest of this work.
In Sec.~\ref{cq}, we canonically quantize the classical field theory associated with
the asymmetric Dirac equation and in Sec.~\ref{ds} we investigate the behavior of 
the asymmetric Dirac fields under the parity, time reversal, and charge conjugation
operations. The Feynman propagator for the asymmetric Dirac fields is computed
in Sec.~\ref{fp} and the interaction of those fields 
with electromagnetic fields is studied in 
Sec.~\ref{qed}. Other types of interactions as well as possible scenarios in which
the asymmetry between particles and antiparticles may lead to beyond standard
model predictions are discussed
in Sec.~\ref{bqed}. Finally, in Sec.~\ref{con} we give our concluding remarks.

\section{The asymmetric Dirac equation}
\label{asd}

\subsection{The wave equation and its main properties}

The asymmetric Dirac equation 
is a relativistic spinorial wave equation that  
has at most first order space-time derivatives and  
that is compatible with the dispersion relations for particles and antiparticles that naturally appeared in the Lorentz covariant 
Schr\"odinger equation \cite{rig22}. This wave equation 
is covariant under Lorentz transformations and a detailed 
discussion of the physical ideas and the mathematical steps leading to 
its derivation can be found in Ref. \cite{rig23}, a prequel to this work.

The asymmetric Dirac equation can be written as \cite{rig23}
\begin{equation}
\boxed{i\hbar \gamma^\mu\partial_\mu\Psi(x) - (imc\gamma^5 - 
\hbar \kappa_\mu \gamma^\mu)\Psi(x) = 0,}
\label{adeFinal2}
\end{equation}
with the particle's mass $m$ given by 
\begin{equation}
m  = \frac{\hbar\sqrt{\kappa^2}}{c}
\label{restmass}
\end{equation}
and
\begin{equation}
\kappa^2=\kappa_\mu\kappa^\mu = (\kappa^0)^2-|\bm{\kappa}|^2
=\widetilde{m}^2=(mc/\hbar)^2,
\label{dispKG}
\end{equation}
where $\bm{\kappa}=(\kappa^1,\kappa^2,\kappa^2)$. Note that we are using 
the Einstein summation convention and the following metric, 
$g_{\mu\nu}=\mbox{diag}\{1,-1,-1,-1\}$. Also, $\gamma^\mu$ denotes the usual 
Dirac gamma matrices and $\gamma^5=i\gamma^0\gamma^1\gamma^2\gamma^3$.

The identification of the mass of the particle as described in Eqs.~(\ref{restmass})
and (\ref{dispKG}) guarantees the empirical equivalence between the standard and
the asymmetric Dirac equations' first quantized predictions \cite{rig23} and,
as we will show here, the equivalence of the second quantized theories at least 
at the level of QED (abelian gauge theories). We should also
note that we only need $\kappa^2=(mc/\hbar)^2$ to establish the equivalence of 
both theories, which means that in general three of the four invariants are 
still free to be adjusted as we wish.

An important particular case occurs when $\kappa^0=mc/\hbar$ and $\kappa^j=0$. In this scenario particles and antiparticles have
different energies for a given wave number while having the same momentum. In general,
and assuming no preferred orientation, we must set $\kappa^1=\kappa^2=\kappa^3$
and $\kappa^0\neq 0$. In this case we are left with only one free parameter after 
satisfying Eq.~(\ref{restmass}) and both the energy and the momentum for particles and
antiparticles sharing the same wave number are no longer degenerate. Note, however,
that by setting $\kappa^1\neq \kappa^2 \neq \kappa^3$ might help us in describing 
anisotropic condensed matter systems \cite{saf93,zha19}.

In Eq.~(\ref{adeFinal2}), $\kappa^\mu$ are four relativistic invariants of the present theory under proper Lorentz transformations (boosts and spatial rotations), while under improper ones they change sign according to the rules ascribed to four-vectors.
We emphasize that $\kappa^\mu$ does not transform as a four-vector under proper Lorentz 
transformations. The parameters $\kappa^0,\kappa^1,\kappa^2$, and $\kappa^3$ are 
the analogs of the rest mass $m$ for the standard Dirac
equation, with $\kappa^\mu$ being the basic invariants of the present theory and 
the mass $m$ a function of them \cite{rig22,rig23}.

The asymmetric Dirac equation was built such that it is covariant under proper and improper Lorentz transformations and such that its four-current $j^\mu(x)$ transforms thereunder 
as a four-vector \cite{rig23}. 
These are the same properties we have for the standard Dirac equation \cite{gre00,man86}. 
The Lagrangian density related to the asymmetric Dirac equation, given in Sec.~\ref{lf}, 
is invariant under those transformations \cite{rig23}. 

For proper Lorentz transformations,  
the wave function $\Psi'(x')$ in the inertial frame $S'$ is connected with 
$\Psi(x)$ in the inertial frame $S$ as follows, 
\begin{equation}
\Psi'(x') = M(x)\Psi(x).
\label{MPsi}
\end{equation}

For an infinitesimal proper Lorentz transformation,
\begin{equation}
x^\mu = x^{\mu'} - \epsilon^{\mu\nu}x_{\nu'}, 
\label{lt}
\end{equation}
$M(x)$ is the following invertible matrix \cite{rig23}, 
\begin{equation}
M(x) = \mathbb{1} + \frac{i}{2}\epsilon_{\mu\nu} (\kappa^\mu x^\nu - \kappa^\nu x^\mu)
- \frac{i}{4}\epsilon_{\mu\nu}\sigma^{\mu\nu}.
\label{Minf}
\end{equation}
Here $\epsilon^{\mu\nu}$ is the infinitesimal antisymmetric real tensor associated with 
the proper Lorentz transformation being implemented \cite{gre00,man86,gre95} and
\begin{equation}
\sigma^{\mu\nu}= (i/2)[\gamma^\mu,\gamma^\nu] = (i/2)(\gamma^\mu\gamma^\nu-
\gamma^\nu\gamma^\mu). 
\end{equation}
We should also note that whenever $\kappa^\mu=0$,
Eq.~(\ref{Minf}) is the transformation 
rule for the standard Dirac spinor.

For finite boosts or finite spatial rotations we have
\cite{rig23}
\begin{equation}
\Psi'(x') = M(x) \Psi(x) = K(x)S\Psi(x).
\label{Mfinite}
\end{equation}
Here $S$ is the corresponding transformation law for a standard Dirac spinor subjected
to the same proper Lorentz transformation under investigation and $K(x)$ is the
corresponding transformation law for a Lorentz-Schr\"odinger  
scalar that satisfies the Lorentz covariant Schr\"odinger
equation \cite{rig22}. The explicit expressions for $K(x)$ are in general given by
$e^{if(x,\kappa)}$, with $f(x,\kappa)$ being real functions of $x^\mu$, $\kappa^\mu$, and
the rapidity (for boosts) or the rotation angle (for spatial rotations) \cite{rig22,rig23}.

\subsection{Lagrangian density and conserved quantities}
\label{lf}

In this section $\Psi(x)$ and $\overline{\Psi}(x)=\Psi^\dagger(x)\gamma^0$ are no longer
treated as ``wave functions''. They are now considered two independent classical spinorial fields which, from Sec.~\ref{cq} on, will be promoted to operators satisfying the standard
canonical anticommutation relations. The Lagrangian that describes the dynamics of those
fields are
\begin{equation}
L = \int d^3x \mathcal{L}[\Psi(x),\overline{\Psi}(x),\partial_\mu \Psi(x),
\partial_\mu \overline{\Psi}(x)],
\label{lag}
\end{equation}
where the Lagrangian density $\mathcal{L}$ depends at most on first order
derivatives of the fields and $d^3x = dx^1dx^2dx^3$ is the
infinitesimal spatial volume. As usual, the integration in Eq.~(\ref{lag})
covers the entire space and the fields and their derivatives are assumed to vanish at the boundaries of integration. Moreover, the dimension of $\mathcal{L}$ is such that   
$L$ has the dimension of energy.

If we define $d^4x = dx^0d^3x=cdtd^3x$ as the infinitesimal four-volume, the action can be
written as \cite{man86,gre95}
\begin{equation}
S = \int dt L = \frac{1}{c} \int d^4x 
\mathcal{L}(\Psi,\overline{\Psi},\partial_\mu \Psi,\partial_\mu \overline{\Psi}),
\label{acao}
\end{equation}
and the principle of least action ($\delta S = 0$) gives the 
Euler-Lagrange equations below \cite{man86,gre95},
\begin{eqnarray}
\frac{\partial \mathcal{L}}{\partial \Psi} = \partial_\mu\left( \frac{\partial \mathcal{L}}{\partial(\partial_\mu \Psi)}\right),\;\;\;
\frac{\partial \mathcal{L}}{\partial \overline{\Psi}} = \partial_\mu\left( \frac{\partial \mathcal{L}}{\partial(\partial_\mu \overline{\Psi})}\right).
\label{euler}
\end{eqnarray}

From Eq.~(\ref{euler}) we obtain the asymmetric Dirac equation 
(\ref{adeFinal2}) and its adjoint
if
\begin{equation}
\boxed{\mathcal{L} = i\hbar c  \overline{\Psi} \gamma^\mu \partial_\mu \Psi
- \overline{\Psi} (imc^2\gamma^5-\hbar c \kappa_\mu\gamma^\mu)  \Psi.}
\label{adeld2}
\end{equation}

Since the Lagrangian density (\ref{adeld2}) is invariant under space-time translations,
spatial rotations, and a global gauge transformation, Noether's theorem gives the following conserved ``charges" \cite{rig23},
\begin{eqnarray}
H &=&  i\hbar \int d^3x 
\Psi^\dagger(x)\partial_t\Psi(x), \label{ham}\\
\mathbf{P} &=& -i\hbar \int d^3x \Psi^\dagger(x)\nabla\Psi(x), \label{p} \\
\mathbf{J} &=&  \int d^3x 
\widetilde{\Psi}^\dagger(x)\left[\mathbf{r}\times \mathbf{\hat{p}}  
+ \frac{\hbar}{2}\bm{\sigma}\right]\widetilde{\Psi}(x), 
\label{jvec2}\\
Q &=& q \int d^3x \Psi^\dagger(x)\Psi(x), 
\label{Q}
\end{eqnarray}
where
\begin{eqnarray}
\mathbf{\hat{p}} &=& -i\hbar\nabla, \\
\bm{\sigma} & = & (\sigma^{23},\sigma^{31},\sigma^{12}), \label{sigma}\\
\widetilde{\Psi}(x) & = & e^{-i\kappa_\mu x^\mu}\Psi(x). \label{aDtoD2}
\end{eqnarray}

Equations (\ref{ham}), (\ref{p}), and (\ref{jvec2}) are, respectively, the energy,
the linear momentum, and the total angular momentum related to the asymmetric Dirac fields.
In Eq.~(\ref{Q}), $q$ will be interpreted in the second quantization formalism as 
the electric charge of a particle and $-q$ as the electric charge of an antiparticle.
The corresponding four-current $j^\mu(x)$, which satisfies the continuity equation 
$\partial_\mu j^\mu(x)=0$, can be written as \cite{rig23}
\begin{equation}
j^\mu(x) = (c\rho(x),\mathbf{j}(x)) = cq\overline{\Psi}(x)\gamma^\mu \Psi(x).
\label{jmu}
\end{equation}

Before we finish this section, we should point out that the asymmetric Dirac Lagrangian
density (\ref{adeld2}) is connected to the standard Dirac Lagrangian density via the
following space-time dependent unitary transformation,
\begin{equation}
\Psi(x) = e^{i\kappa_\mu x^\mu}U\Psi_D(x),
\label{aDtoD}
\end{equation}
where
\begin{equation}
U = \left(\frac{\mathbb{1}-i\gamma^5}{\sqrt{2}}
\right). \label{U}
\end{equation}
In Eq.~(\ref{aDtoD}), the subscript ``D'' denotes the standard Dirac field.
If we insert Eq.~(\ref{aDtoD}) into (\ref{adeld2}) we get,
\begin{equation}
\mathcal{L}_D = i\hbar c  \overline{\Psi}_D \gamma^\mu \partial_\mu \Psi_D
- mc^2\overline{\Psi}_D\Psi_D,
\end{equation}
which is exactly the standard Dirac Lagrangian density.

\section{Canonical quantization}
\label{cq}

\subsection{Plane wave solutions}

The positive and negative plane wave solutions to the asymmetric Dirac equation are
proportional to, respectively \cite{rig23},
\begin{eqnarray}
u_r(\mathbf{p})e^{i\kappa_\mu x^\mu} e^{-ip_\mu x^\mu/\hbar}, \label{pwu} \\
v_r(\mathbf{p})e^{i\kappa_\mu x^\mu} e^{ip_\mu x^\mu/\hbar}, \label{pwv}
\end{eqnarray}
where $u_r(\mathbf{p}), v_r(\mathbf{p})$ are a set of four linearly independent spinors
($r=1,2$) that satisfy the following equations,
\begin{eqnarray}
(\slashed p - i m c \gamma^5)u_r(\mathbf{p}) &=& 0, 
\label{ue} \\
(\slashed p + i m c \gamma^5)v_r(\mathbf{p}) &=& 0, \label{ve}
\end{eqnarray}
with $\slashed p = \gamma^\mu p_\mu$ being Feynman slash notation.
Note that $u_r(\mathbf{p})$ and $v_r(\mathbf{p})$ do not depend on $\kappa^\mu$ and we 
also have that \cite{rig23}
\begin{equation}
p^\mu = \hbar k^\mu \label{pmu}
\end{equation}
is a four-vector satisfying the usual energy-momentum relations for standard relativistic
particles,
\begin{eqnarray}
p^\mu p_\mu &=& m^2c^2 , \label{psquared}\\
p^0 &=& E_{\mathbf{p}}/c, \label{p0}\\
E_{\mathbf{p}} & = & \sqrt{m^2c^4 + |\mathbf{p}|^2c^2}. \label{Ep}
\end{eqnarray}
Also, the vector $\mathbf{k}=(k^1,k^2,k^3)$ is the particle's wave vector and we 
once more emphasize that $\kappa^\mu$ is not a four-vector, being the four relativistic
invariants of the present theory.

Since the helicity projection operators commute with the energy projection
operators \cite{rig23}, we follow the standard convention 
and choose $u_r(\mathbf{p}), v_r(\mathbf{p})$ such that \cite{man86}

\begin{eqnarray}
\sigma_{\mathbf{p}}u_r(\mathbf{p})&=&(-1)^{r+1}u_r(\mathbf{p}), \label{hu} \\
\sigma_{\mathbf{p}}v_r(\mathbf{p})&=&(-1)^{r}v_r(\mathbf{p}), \label{hv}
\end{eqnarray}
where
\begin{equation}
\sigma_{\mathbf{p}} = \frac{\bm{\sigma}\cdot \mathbf{p}}{|\mathbf{p}|},
\label{sigmap}
\end{equation}
with $\bm{\sigma}$ given by Eq.~(\ref{sigma}). 

Working with the appropriate normalization \cite{gre00,man86}, the spinors 
satisfy the 
following orthogonality relations \cite{rig23},
\begin{eqnarray}
u_r^\dagger(\mathbf{p})u_s(\mathbf{p}) &=& v_r^\dagger(\mathbf{p})v_s(\mathbf{p}) 
= 
\frac{E_{\mathbf{p}}}{mc^2}\delta_{rs}=\frac{p^0}{mc}\delta_{rs},
\label{urvs} \\
u_r^\dagger(\mathbf{p})v_s(\mathbf{-p}) &=& v_r^\dagger(\mathbf{p})u_s(\mathbf{-p}) 
=  0, \label{o1}
\end{eqnarray}
where $\delta_{rs}$ is the Kronecker delta.

\subsection{Second quantization}

The first step to second quantize the asymmetric Dirac fields is to expand them in terms of 
the complete set of plane waves given by Eqs.~(\ref{pwu}) and (\ref{pwv}),
\begin{eqnarray}
\Psi(x) &=& \Psi^+(x) + \Psi^-(x), \label{psi}\\ 
\overline{\Psi}(x) &=& \overline{\Psi}^+(x) + \overline{\Psi}^-(x), \label{psibar} 
\end{eqnarray}
where
\begin{eqnarray}
\Psi^+(x) & = & \sum_{r=1}^2\int \widetilde{dp}c_r(\mathbf{p})u_r(\mathbf{p})
e^{i\kappa x}e^{-ipx/\hbar}, \label{psi+}\\ 
\Psi^-(x) & = & \sum_{r=1}^2\int \widetilde{dp}d^\dagger_r(\mathbf{p})v_r(\mathbf{p})
e^{i\kappa x}e^{ipx/\hbar}, \label{psi-}\\
\overline{\Psi}^+(x) & = & \sum_{r=1}^2\int \widetilde{dp}d_r(\mathbf{p})
\overline{v}_r(\mathbf{p})
e^{-i\kappa x}e^{-ipx/\hbar}, \label{psibar+}\\ 
\overline{\Psi}^-(x) & = & \sum_{r=1}^2\int \widetilde{dp}c^\dagger_r(\mathbf{p}) \overline{u}_r(\mathbf{p})e^{-i\kappa x}e^{ipx/\hbar}, \label{psibar-}
\end{eqnarray}
and
\begin{eqnarray}
\widetilde{dp} &=& g(|\mathbf{p}|)d^3p, \label{widedp} \\
g(|\mathbf{p}|) &=& \sqrt{\frac{mc^2}{(2\pi\hbar)^3E_{\mathbf{p}}}}.
\label{gp}
\end{eqnarray}
Here $p^\mu$ and $E_{\mathbf{p}}$ are such that they satisfy Eqs.~(\ref{pmu})-(\ref{Ep}) and
$\overline{w}_r=w^\dagger_r\gamma^0$, with $w=u,v$. Since we are adopting the
infinite volume normalization for the plane waves \cite{gre95}, 
the integrals run from $-\infty$ to $\infty$ and we have $g(|\mathbf{p}|)$ 
as given above instead of the finite volume normalization given in Ref.~\cite{man86}.

The second step towards the canonical quantization of the asymmetric Dirac fields is 
the identification of the expansion coefficients $c_r(\mathbf{p}), d_r(\mathbf{p})$
as annihilation operators and $c^\dagger_r(\mathbf{p}), d^\dagger_r(\mathbf{p})$
as creation operators satisfying the following anticommutation relations,
\begin{equation}
\{c_r(\mathbf{p}),c_s^\dagger(\mathbf{p'}) \} 
= \{d_r(\mathbf{p}),d_s^\dagger(\mathbf{p'}) \}
= \delta_{rs}\delta^{(3)}(\mathbf{p}-\mathbf{p'}), \label{aad}
\end{equation}
where $\delta^{(3)}(\mathbf{p}-\mathbf{p'})=\delta(p^1- p^{1'})
\delta(p^2- p^{2'})\delta(p^3-p^{3'})$ is the three-dimensional Dirac delta function.
Any other anticommutator involving two of those four operators is zero. 

Now, looking at the plane wave expansions given by Eqs.~(\ref{psi+})-(\ref{psibar-}),
we realize that they are almost equal to the ones employed in the canonical quantization
of the standard Dirac fields \cite{man86,gre95}. 
The only differences are the exponentials $e^{\pm i\kappa x}$. Therefore, almost all 
textbook calculations related to the canonical quantization of the standard Dirac field
can be carried over to the ones we will be doing here if we replace 
$p$ with $p-\hbar\kappa$ in Eqs.~(\ref{psi+}) and (\ref{psibar-}) and 
$p$ with $p+\hbar\kappa$ in Eqs.~(\ref{psi-}) and (\ref{psibar+}). With that in mind,
a direct calculation after inserting Eqs.~(\ref{psi}) and (\ref{psibar}) into 
Eqs.~(\ref{ham}), (\ref{p}), and (\ref{Q}) gives,
\begin{eqnarray}
:H: &=&  \sum_{r=1}^2\int d^3p [(E_{\mathbf{p}}-\hbar c \kappa^0)
c^\dagger_r(\mathbf{p})c_r(\mathbf{p}) \nonumber \\
&&+(E_{\mathbf{p}}+\hbar c \kappa^0)d^\dagger_r(\mathbf{p})d_r(\mathbf{p})], \label{ham2}\\
:\mathbf{P}: &=& \sum_{r=1}^2\int d^3p [(\mathbf{p}-\hbar \bm{\kappa})
c^\dagger_r(\mathbf{p})c_r(\mathbf{p}) \nonumber \\
&&+(\mathbf{p}+\hbar \bm{\kappa})d^\dagger_r(\mathbf{p})d_r(\mathbf{p})], \label{p2} \\
:Q: &=& q\sum_{r=1}^2\int d^3p [c^\dagger_r(\mathbf{p})c_r(\mathbf{p}) -
d^\dagger_r(\mathbf{p})d_r(\mathbf{p})]. \label{Q2}
\end{eqnarray}
To arrive at Eqs.~(\ref{ham2})-(\ref{Q2}), we used the orthogonality relations 
(\ref{urvs}) and (\ref{o1}), the following representation for the Dirac delta,
\begin{equation}
\delta^{(3)}(\mathbf{p-p'}) = \frac{1}{(2\pi\hbar)^3}\int d^3x \;
e^{\pm i(\mathbf{p-p'})\cdot 
\mathbf{r}/\hbar}, \label{diracdelta}
\end{equation}
and the following identities \cite{rig22},
\begin{eqnarray}
\frac{1}{(2\pi\hbar)^3}\!\int\!\! d^3x\; e^{\pm i (p-p')x/\hbar} \!\!&=&\!\!
\delta^{(3)}(\mathbf{p-p'}), \label{diracM}\\
\frac{1}{(2\pi\hbar)^3}\!\int\!\! d^3x\; e^{\pm i (p+p')x/\hbar} \!\!&=&\!\! 
e^{\pm i(p^0+p^{0'})x^0/\hbar} \delta^{(3)}(\mathbf{p+p'}), \nonumber \\
& &  \label{diracP} 
\end{eqnarray}
where $p^0$ and $p^{0'}$ are given by Eq.~(\ref{p0}).
Note also that in Eqs.~(\ref{ham2})-(\ref{Q2}), the symbol ``$:\;:$'' denotes 
the standard fermion normal ordering prescription \cite{man86,gre95}.
And similarly to the standard Dirac fields, the quantization of the asymmetric 
Dirac fields using fermionic statistics was crucial to arrive at Eq.~(\ref{ham2}), 
where the energies of particles and antiparticles are bounded from below. Had we used 
bosonic statistics, the normal ordering prescription would have led to antiparticles
with energies given by $-(E_{\mathbf{p}}+\hbar c \kappa^0)$, which have no lower bounds 
and tend to $-\infty$ when $|\mathbf{p}|\rightarrow \infty$.

\subsubsection*{The asymmetry between particles and antiparticles}

The first remarkable feature of the asymmetric Dirac fields can be seen at
Eqs.~(\ref{ham2}) and (\ref{p2}), which justifies why we called the present free field
theory ``asymmetric''. Particles and antiparticles no longer have the same energy and
momentum for a given wave number. The energy for a single vacuum excitation (a particle)
with wave number $\mathbf{k}=\mathbf{p}/\hbar$ is 
$E_{\mathbf{p}}^-=E_{\mathbf{p}}-\hbar c \kappa^0$ while an antiparticle has 
an energy given by 
$E_{\mathbf{p}}^+=E_{\mathbf{p}}+\hbar c \kappa^0$. Their momenta are, respectively,
$\mathbf{p}_-=\mathbf{p}-\hbar \bm{\kappa}$ and 
$\mathbf{p}_+=\mathbf{p}+\hbar \bm{\kappa}$. And of course, what we call particles and
antiparticles are a mere convention. We chose the present one since for positive $\kappa^0$,
particles are less energetic than antiparticles, which might help in the understanding 
of why particles dominate antiparticles in our universe \cite{rig22}.

As briefly discussed in Sec.~\ref{asd}, an important particular case occurs when we fix 
$\kappa^0=mc/\hbar$ and $\kappa^j=0$. In this scenario, 
$E_{\mathbf{p}}^{\mp}=E_{\mathbf{p}} \mp mc^2$ and 
$\mathbf{p}_{\mp}=\mathbf{p}$. In other words, only the energies of particles and 
antiparticles differ while their momenta are equal and identical to the momenta of
standard Dirac particles and antiparticles. For this particular choice of parameters,
the gap between the energy of an antiparticle and a particle sharing the same wave number
is $2mc^2$. 

The other interesting case happens when we assume isotropy of space, as before,
but with $\bm{\kappa}\neq 0$. In this case $\kappa^1=\kappa^2=\kappa^3$ and 
$\kappa^0\neq 0$ such that $\kappa^2=(mc/\hbar)^2$. Now both the energy and the momentum
are no longer degenerate. The gap between the energy of an antiparticle and a particle 
having the same wave number is $2\hbar c \kappa^0$. Although the gap in the energies
between particles and antiparticles
suggests a possible way to understand the asymmetry 
between matter and antimatter in the present day universe \cite{rig22}, a physical meaning 
to the gap in momentum is not completely understood \cite{rig22,rig23}. 

Nevertheless, as we showed for the scalar field theory associated with the Lorentz covariant 
Schr\"odinger equation in Ref. \cite{rig22}, and as we will show for the 
spin-1/2 quantum field theory developed here,
we can make those theories equivalent to the Klein-Gordon and Dirac field theories, 
respectively, as long as we guarantee that $\kappa^2=(mc/\hbar)^2$ and 
whether or not we have $\bm{\kappa}= 0$. This equivalence is true 
for the free field case,  when we include self interactions respecting the 
Lorentz symmetry, and for interactions mediated by massless abelian gauge fields 
such as QED. 
Further investigations are needed to check what happens for interactions mediated by
non-abelian gauge fields, in particular when we have massive gauge bosons 
described by fields that transform under proper Lorentz transformations 
analogously to the generalized transformation rules 
we introduced for scalar and spinorial fields in Refs.~\cite{rig22,rig23}. 

When it comes to the angular momentum, orbital or intrinsic,
the asymmetry is not present. For those quantities, particles and antiparticles are still
symmetric. This can be understood looking at Eq.~(\ref{jvec2}), which does not depend on
$\kappa^\mu$ since $\widetilde{\Psi}(x)$ and $\widetilde{\Psi}^\dagger(x)$ do 
not depend on it. Indeed, the exponential
$e^{-ikx}$ that appears in the definition of 
$\widetilde{\Psi}(x)$  [cf. Eq.~(\ref{aDtoD2})] 
cancels the exponential $e^{ikx}$ appearing in the plane wave expansion of 
$\Psi(x)$ [cf. Eqs.~(\ref{psi}), (\ref{psi+}), and (\ref{psi-})]. A similar argument
applies to prove that $\widetilde{\Psi}^\dagger(x)$ does not depend on $\kappa^\mu$ either.

We should also note that according to Eqs.~(\ref{jvec2}) and (\ref{sigmap}), 
we can define the helicity operator for the asymmetric Dirac field, i.e., its intrinsic angular momentum projected in the direction of the wave number 
$\mathbf{k}=\mathbf{p}/\hbar$, as follows \cite{man86},
\begin{equation}
S_{\mathbf{p}} =  \frac{\hbar}{2}\int d^3x 
:\widetilde{\Psi}^\dagger(x)\sigma_{\mathbf{p}}\widetilde{\Psi}(x):. 
\label{sp}
\end{equation}
A direct calculation similar to the ones that led to Eqs.~(\ref{ham2})-(\ref{Q2}) gives
\begin{equation}
S_{\mathbf{p}} =  \frac{\hbar}{2}\sum_{r=1}^2\int d^3p (-1)^{r+1}[c^\dagger_r(\mathbf{p})c_r(\mathbf{p}) +
d^\dagger_r(\mathbf{p})d_r(\mathbf{p})]. \label{sp2}
\end{equation}
To arrive at Eq.~(\ref{sp2}) we also 
made use Eqs.~(\ref{hu}) and (\ref{hv}). Note that
the $(-1)^r$ appearing in Eq.~(\ref{hv}) is multiplied by the $-1$ 
that we get when applying the normal ordering prescription to write 
$d^\dagger_r(\mathbf{p})$ to the left of $d_r(\mathbf{p})$. This is why we have
an overall $(-1)^{r+1}$ in Eq.~(\ref{sp2}).

If we use the anticommutation relations given by Eq.~(\ref{aad}), it is not difficult 
to see that Eq.~(\ref{sp2}) gives
\begin{eqnarray}
S_{\mathbf{p}} c^\dagger_r(\mathbf{p}) |0\rangle &=& (-1)^{r+1}\frac{\hbar}{2}
c^\dagger_r(\mathbf{p}) |0\rangle, \label{spinc} \\
S_{\mathbf{p}} d^\dagger_r(\mathbf{p}) |0\rangle &=& (-1)^{r+1}\frac{\hbar}{2}
d^\dagger_r(\mathbf{p}) |0\rangle, \label{spind}
\end{eqnarray}
where $|0\rangle$ is the vacuum state, 
\begin{equation}
c_r(\mathbf{p}) |0\rangle = d_r(\mathbf{p}) |0\rangle = 0.
\end{equation}

Equations (\ref{ham2})-(\ref{Q2}) together with (\ref{spinc})-(\ref{spind}), fully justify
the interpretation of the vacuum excitation created by $c_r^\dagger(\mathbf{p})$ 
as a particle with
energy $E_{\mathbf{p}}-\hbar c \kappa^0$, momentum $\mathbf{p}-\hbar\bm{\kappa}$, 
charge $q$, and spin $(-1)^{r+1}\hbar/2$ along the direction of its wave number, while
the vacuum excitation created by $d_r^\dagger(\mathbf{p})$ is interpreted 
as an antiparticle with
energy $E_{\mathbf{p}}+\hbar c \kappa^0$, momentum $\mathbf{p}+\hbar\bm{\kappa}$, 
charge $-q$, and spin $(-1)^{r+1}\hbar/2$ along the direction of its wave number.

\section{Discrete symmetries}
\label{ds}

Our goal now is to study in the second quantization context the discrete symmetry 
operations investigated in Ref.~\cite{rig23} when we developed the first quantized 
theory of the asymmetric Dirac equation. We will show in what follows that 
we can define the parity, time reversal, and charge conjugation operations 
in such a way that we have the Lagrangian 
density (\ref{adeld2}) transforming under those symmetry operations 
in exactly the same way the standard Dirac Lagrangian density does.
From now on it is implicit that the normal ordering prescription is applied. 
In particular, when we write or refer to Eqs.~(\ref{adeld2})-(\ref{jvec2}),
we always mean the normal ordered expressions.

\subsection{Space inversion or parity}

The parity operation changes the sign of the space
coordinates associated with a physical system,
\begin{equation}
x = (x^0, \mathbf{r}) \longrightarrow x' = (x^{0'}, \mathbf{r'}) = (x^0,-\mathbf{r}).
\label{si}
\end{equation}

Building on the first quantized parity operator of Ref.~\cite{rig23}, the 
second quantized parity operator transforms an asymmetric Dirac field as follows, 
\begin{equation}
\mathcal{P}_{\kappa} \Psi(x^0,\mathbf{r})\mathcal{P}_{\kappa}^\dagger =
\gamma^5P\Psi(x^0,-\mathbf{r}),
\label{parityOp}
\end{equation}
where
\begin{eqnarray}
\mathcal{P}_{\kappa} &=& K_1 \mathcal{P}, \label{pk1}\\
P &=& e^{i\varphi_{\!_P}}\gamma^0,\\
K_1 f(\kappa^\mu) K_1^\dagger &=& f(\kappa_\mu).
\label{k1}
\end{eqnarray}
%
The unitary operator 
$\mathcal{P}$ lives on the Fock space and thus it affects the creation and annihilation operators while the unitary operator $P$ acts on spinors.
Also, $f(\kappa^\mu)$ is an arbitrary function of 
$\kappa^\mu$, $K_1 = K_1^\dagger$, and $(K_1)^2=1$ \cite{rig22}. In the 
metric signature we use here, $K_1$ changes the sign of $\kappa^j$ and leaves
$\kappa^0$ unchanged. Note that the sign of the mass $m$ is not altered 
by $K_1$ since $m$ depends quadratically on $\kappa^\mu$. Also,
$e^{i\varphi_{\!_P}}=\pm 1$ or $\pm i$ if we postulate that four successive 
applications of the parity operator should bring us back to where we started
\cite{gre00,gre95}. We should also mention that a standard Dirac field satisfies
an equation similar to Eq.~(\ref{parityOp}),
with $\mathcal{P}_{\kappa}$ changed to $\mathcal{P}$ 
and a right hand side without the $\gamma^5$.

The parity operator defined by Eqs.~(\ref{parityOp})-(\ref{k1}) is such that 
the asymmetric Dirac equation is covariant under its action. Moreover, a direct
calculation shows that \cite{rig23}
\begin{eqnarray}
\mathcal{P}_{\kappa}\mathcal{L}(x^0,\mathbf{r})\mathcal{P}_{\kappa}^\dagger &=& \mathcal{L}(x^0,\mathbf{-r}), \label{parity1}\\
\mathcal{P}_{\kappa}H\mathcal{P}_{\kappa}^\dagger &=& 
H, \\
\mathcal{P}_{\kappa}\mathbf{P}\mathcal{P}_{\kappa}^\dagger &=& 
-\mathbf{P}, \\
\mathcal{P}_{\kappa}\mathbf{J}\mathcal{P}_{\kappa}^\dagger &=& 
\mathbf{J}, \label{parity4}
\end{eqnarray}
where the Lagrangian density, the Hamiltonian, the linear momentum,
and the total angular momentum are given by the second quantized 
version of Eqs.~(\ref{adeld2})-(\ref{jvec2}). These are the expected 
transformation laws for those quantities and exactly the same transformation 
laws we have for standard Dirac fields. 

Using Eq.~(\ref{parityOp}) we can obtain how the creation and annihilation operators 
change after the parity operation. Inserting Eq.~(\ref{psi}) into the left hand side
of Eq.~(\ref{parityOp}) we get after using Eq.~(\ref{k1}),
\begin{eqnarray}
\mathcal{P}_{\kappa} \Psi(x^0,\mathbf{r})\mathcal{P}_{\kappa}^\dagger &=&
\sum_{r=1}^2\int \widetilde{dp}
[\mathcal{P}_{\kappa}c_r(\mathbf{p})\mathcal{P}_{\kappa}^\dagger]    
u_r(\mathbf{p})e^{i\widetilde{\kappa} x}e^{-ipx/\hbar} \nonumber \\
& + & \sum_{r=1}^2\int \widetilde{dp}
[\mathcal{P}_{\kappa}d^\dagger_r(\mathbf{p})\mathcal{P}_{\kappa}^\dagger] 
v_r(\mathbf{p})e^{i\widetilde{\kappa} x}e^{ipx/\hbar}, \nonumber \\
& & \label{leparity}
\end{eqnarray}
where 
\begin{equation}
\widetilde{\kappa}^\mu=(\kappa^0,-\bm{\kappa}). 
\end{equation}
On the other hand,
inserting Eq.~(\ref{psi}) into the right hand side of Eq.~(\ref{parityOp}),
using that \cite{rig23}
\begin{eqnarray}
u_r(\mathbf{p}) &=&  i\gamma^0\gamma^5u_r(\mathbf{-p}),  \label{i3}\\
v_r(\mathbf{p}) &=& - i\gamma^0\gamma^5v_r(\mathbf{-p}), \label{i4}
\end{eqnarray}
we obtain after a change of variables ($\mathbf{p}\rightarrow -\mathbf{p}$),
\begin{eqnarray}
\gamma^5P\Psi(x^0,-\mathbf{r}) \hspace{-.2cm}&=&\hspace{-.2cm}
\sum_{r=1}^2\int \widetilde{dp}
[ie^{i\varphi_{\!_P}}c_r(\mathbf{-p})]    
u_r(\mathbf{p})e^{i\widetilde{\kappa} x}e^{-ipx/\hbar} \nonumber \\
\hspace{-.2cm}& + &\hspace{-.2cm} \sum_{r=1}^2\int \widetilde{dp}
[-ie^{i\varphi_{\!_P}}d^\dagger_r(\mathbf{-p})] 
v_r(\mathbf{p})e^{i\widetilde{\kappa} x}e^{ipx/\hbar}, \nonumber \\
& & \label{ldparity}
\end{eqnarray}

Since the right hand side of Eq.~(\ref{parityOp})
must be equal to its left hand side, comparing Eqs.~(\ref{leparity}) and (\ref{ldparity})
we obtain the transformation rules for the creation and annihilation operators,
\begin{eqnarray}
\mathcal{P}_{\kappa}c_r(\mathbf{p})\mathcal{P}_{\kappa}^\dagger &=&
ie^{i\varphi_{\!_P}}c_r(\mathbf{-p}), \label{cp}\\
\mathcal{P}_{\kappa}d_r^\dagger(\mathbf{p})\mathcal{P}_{\kappa}^\dagger &=&
-ie^{i\varphi_{\!_P}} d_r^\dagger(\mathbf{-p}). \label{ddaggerp}
\end{eqnarray}

If we choose $\varphi_{\!_P}=-\pi/2$, we recover the usual transformation rules 
associated with the standard Dirac fields \cite{gre95},
\begin{eqnarray}
\mathcal{P}_{\kappa}c_r(\mathbf{p})\mathcal{P}_{\kappa}^\dagger =
c_r(\mathbf{-p});& 
\mathcal{P}_{\kappa}c_r^\dagger(\mathbf{p})\mathcal{P}_{\kappa}^\dagger 
= c_r^\dagger(\mathbf{-p}), \\
\mathcal{P}_{\kappa}d_r(\mathbf{p})\mathcal{P}_{\kappa}^\dagger =
- d_r(\mathbf{-p});&
\mathcal{P}_{\kappa}d_r^\dagger(\mathbf{p})\mathcal{P}_{\kappa}^\dagger =
- d_r^\dagger(\mathbf{-p}). \label{dddagger}
\end{eqnarray}
The minus sign in Eq.~(\ref{dddagger}) means that particles and 
antiparticles have opposite intrinsic parity, similarly to standard Dirac particles and
antiparticles. 

\textit{Remark.} In the calculations that led to
Eqs.~(\ref{cp}) and (\ref{ddaggerp}), it is crucial
to define $\mathcal{P}_{\kappa}$ as given by Eq.~(\ref{pk1}). We need the operator 
$K_1$ to change the sign of $\bm{\kappa}$ in the left hand side. This gives
$e^{i\widetilde{\kappa}x}$, which matches exactly the $e^{i\widetilde{\kappa}x}$ 
in the right hand side. As we will see next, similar arguments apply to the time reversal
and charge conjugation operations too.

\subsection{Time reversal}

The time reversal operation changes the sign of the time
coordinate,
\begin{equation}
x = (x^0, \mathbf{r}) \longrightarrow x' = (x^{0'}, \mathbf{r'}) = (-x^0,\mathbf{r}).
\label{tr}
\end{equation}

According to Ref.~\cite{rig23}, the
second quantized time reversal operator should be defined as, 
\begin{equation}
\mathcal{T}_{\kappa} \Psi(x^0,\mathbf{r})\mathcal{T}_{\kappa}^\dagger =
\gamma^5T_0\Psi(-x^0,\mathbf{r}),
\label{timereversalOp}
\end{equation}
where
\begin{eqnarray}
\mathcal{T}_{\kappa} &=& K_1 \mathcal{T}, \label{tk1}\\
T_0 &=& e^{i\varphi_{\!_T}}i\gamma^1\gamma^3.
\end{eqnarray}
%
The unitary operator $T_0$ above, 
that acts on the spinor space, 
has that particular form in the Dirac-Pauli representation for the gamma matrices \cite{gre00,gre95}.  
Also, $\varphi_{\!_T}$ is a real number and we should not forget to mention 
that $\mathcal{T}$,
and consequently $\mathcal{T}_{\kappa}$, are antiunitary operators, namely,
$\mathcal{T} z \mathcal{T}^\dagger =z^*$, 
where $z$ is a complex number and $z^*$ its
complex conjugate. Note also that the equivalent transformation rule for a 
standard Dirac field is given by Eq.~(\ref{timereversalOp}) with 
$\mathcal{T}_{\kappa}$ changed to $\mathcal{T}$ and a 
right hand side without the $\gamma^5$.

The asymmetric Dirac equation is covariant after the action of the time reversal operator defined by Eq.~(\ref{timereversalOp}) and a direct
computation gives \cite{rig23}
\begin{eqnarray}
\mathcal{T}_{\kappa}\mathcal{L}(x^0,\mathbf{r})\mathcal{T}_{\kappa}^\dagger &=& 
\mathcal{L}(-x^0,\mathbf{r}), \label{time1}\\
\mathcal{T}_{\kappa}H\mathcal{T}_{\kappa}^\dagger &=& 
H, \\
\mathcal{T}_{\kappa}\mathbf{P}\mathcal{T}_{\kappa}^\dagger &=& 
-\mathbf{P}, \\
\mathcal{T}_{\kappa}\mathbf{J}(x^0)\mathcal{T}_{\kappa}^\dagger &=& 
-\mathbf{J}(-x^0). \label{time4}
\end{eqnarray}
These are the expected 
transformation laws for Eqs.~(\ref{adeld2})-(\ref{jvec2}) after the time 
reversal operation and the ones we obtain working with the usual Dirac fields
\cite{gre95}. 

As we show next, the creation and annihilation operators transformation laws after 
the time reversal operation is a bit more complex than the analog expressions 
for the parity operation. Inserting Eq.~(\ref{psi}) into the left hand side
of Eq.~(\ref{timereversalOp}), using Eq.~(\ref{k1}), and the antiunitarity 
of $\mathcal{T}_{\kappa}$, we get
\begin{eqnarray}
\mathcal{T}_{\kappa} \Psi(x^0,\mathbf{r})\mathcal{T}_{\kappa}^\dagger &=&
\sum_{r=1}^2\int \widetilde{dp}
[\mathcal{T}_{\kappa}c_r(\mathbf{p})\mathcal{T}_{\kappa}^\dagger]    
u^*_r(\mathbf{p})e^{-i\widetilde{\kappa} x}e^{ipx/\hbar} \nonumber \\
& + &\!\! \sum_{r=1}^2\int \widetilde{dp}
[\mathcal{T}_{\kappa}d^\dagger_r(\mathbf{p})\mathcal{T}_{\kappa}^\dagger] 
v^*_r(\mathbf{p})e^{-i\widetilde{\kappa} x}e^{-ipx/\hbar}. \nonumber \\
& & \label{letimereversal}
\end{eqnarray}

Now, inserting Eq.~(\ref{psi}) into the right hand side of Eq.~(\ref{timereversalOp})
we get
\begin{eqnarray}
\gamma^5T_0\Psi(-x^0,\mathbf{r}) \hspace{-.2cm}&=&\hspace{-.2cm}
\sum_{r=1}^2\int \widetilde{dp}
[e^{i\varphi_{\!_T}}(-1)^{r+1}c_{\overline{r}}(\mathbf{-p})] \nonumber \\    
&&\hspace{1cm}\times u^*_r(\mathbf{p})e^{-i\widetilde{\kappa} x}e^{ipx/\hbar} \nonumber \\
\hspace{-.2cm}& + &\hspace{-.2cm} \sum_{r=1}^2\int \widetilde{dp}
[e^{i\varphi_{\!_T}}(-1)^{r+1}d^\dagger_{\overline{r}}(\mathbf{-p})] \nonumber \\ 
&&\hspace{1cm}\times v^*_r(\mathbf{p})e^{-i\widetilde{\kappa} x}e^{-ipx/\hbar}, \nonumber \\
& & \label{ldtimereversal}
\end{eqnarray}
where
\begin{equation}
\overline{r} = 
\left\{
\begin{array}{ll}
1, \;\mbox{if}\;\; r=2, \\
2, \;\mbox{if}\;\; r=1.
\end{array}
\right.
\label{rbar2}
\end{equation}

To arrive at Eq.~(\ref{ldtimereversal}) we relabeled the summation index 
($r \rightarrow \overline{r}$), used that $(-1)^{\overline{r}}=(-1)^{r+1}$, and used
the following identity, valid for the Dirac-Pauli representation of the gamma matrices,
\begin{equation}
i\gamma^5\gamma^1\gamma^3w_r(-\mathbf{p})=(-1)^rw^*_{\overline{r}}(\mathbf{p}),
\label{identityg5}
\end{equation}
where $w=u,v$. To prove Eq.~(\ref{identityg5}), we use the expressions for 
$u_r(\mathbf{p})$ and $v_r(\mathbf{p})$ in the Dirac-Pauli representation given in 
Ref.~\cite{rig23} and show by an explicit calculation that Eq.~(\ref{identityg5}) 
is true.

Comparing Eqs.~(\ref{letimereversal}) and (\ref{ldtimereversal}), the left and 
right hand sides of Eq.~(\ref{timereversalOp}), we realize that 
we must have,
\begin{eqnarray}
\mathcal{T}_{\kappa}c_r(\mathbf{p})\mathcal{T}_{\kappa}^\dagger &=&
e^{i\varphi_{\!_T}}(-1)^{r+1}c_{\overline{r}}(\mathbf{-p}), \label{ct}\\
\mathcal{T}_{\kappa}d_r^\dagger(\mathbf{p})\mathcal{T}_{\kappa}^\dagger &=&
e^{i\varphi_{\!_T}}(-1)^{r+1} d_{\overline{r}}^\dagger(\mathbf{-p}). \label{ddaggert}
\end{eqnarray}

If we choose $\varphi_{\!_T}=\pi/2$, we obtain the transformation rules 
associated with the standard Dirac fields \cite{gre95},
\begin{eqnarray}
\mathcal{T}_{\kappa}c_r(\mathbf{p})\mathcal{T}_{\kappa}^\dagger &=&i(-1)^{r+1}
c_{\overline{r}}(\mathbf{-p}), \\ 
\mathcal{T}_{\kappa}c_r^\dagger(\mathbf{p})\mathcal{T}_{\kappa}^\dagger 
&=& -i(-1)^{r+1} c_{\overline{r}}^\dagger(\mathbf{-p}), \\
\mathcal{T}_{\kappa}d_r(\mathbf{p})\mathcal{T}_{\kappa}^\dagger &=&-i(-1)^{r+1}
d_{\overline{r}}(\mathbf{-p}), \\
\mathcal{T}_{\kappa}d_r^\dagger(\mathbf{p})\mathcal{T}_{\kappa}^\dagger &=&i(-1)^{r+1}
d_{\overline{r}}^\dagger(\mathbf{-p}). 
\end{eqnarray}

Important relations that are corollaries to Eqs.~(\ref{ct}) and (\ref{ddaggert}) are 
\begin{eqnarray}
\mathcal{T}_{\kappa}c_r^\dagger(\mathbf{p})c_r(\mathbf{p})\mathcal{T}_{\kappa}^\dagger
&=&c_{\overline{r}}^\dagger(\mathbf{-p})c_{\overline{r}}(\mathbf{-p}), \\
\mathcal{T}_{\kappa}d_r^\dagger(\mathbf{p})d_r(\mathbf{p})\mathcal{T}_{\kappa}^\dagger
&=&d_{\overline{r}}^\dagger(\mathbf{-p})d_{\overline{r}}(\mathbf{-p}),
\end{eqnarray}
which help in proving Eqs.~(\ref{time1})-(\ref{time4}) if we work with 
Eqs.~(\ref{adeld2})-(\ref{jvec2}) written in terms of the creation and annihilation
operators.

\subsection{Charge conjugation}

The second quantized version of the charge conjugation operation given in Ref.~\cite{rig23} 
is
\begin{equation}
\mathcal{C}_{\kappa} \Psi(x)\mathcal{C}_{\kappa}^\dagger =
C\overline{\Psi}^T(x),
\label{chargeOp}
\end{equation}
where
\begin{eqnarray}
\mathcal{C}_{\kappa} &=& K_2 \mathcal{C}, \label{ck2}\\
C &=& e^{i\varphi_{\!_C}}i\gamma^2\gamma^0, \label{c} \\
K_2 f(\kappa^\mu) K_2^\dagger &=& f(-\kappa^\mu).
\label{k2}
\end{eqnarray}
Here $\mathcal{C}$ and $C$ are 
unitary operators and $T$ denotes the transpose operation. 
The expression for $C$, which acts in the spinor space, 
is valid in the Dirac-Pauli representation for the gamma matrices \cite{gre00,gre95}. Moreover, $\varphi_{\!_C}$ is a real number, 
$K_2 = K_2^\dagger$, and $(K_2)^2=1$ \cite{rig22}. Note that 
$K_2$ changes the sign of $\kappa^\mu$, leaving the mass $m$ unaltered 
since the latter depends quadratically on $\kappa^\mu$. For a standard 
Dirac spinor, we have the same transformation rule given above, with 
$\mathcal{C}_{\kappa}$ replaced by $\mathcal{C}$.

Similarly to the other two discrete symmetries studied above, 
the asymmetric Dirac equation is covariant under the action 
of the charge conjugation
operator as given by Eq.~(\ref{chargeOp}). Also, employing the same 
techniques given in Ref. \cite{rig23}, it is not difficult to see that at the 
second quantization level we have
\begin{eqnarray}
\mathcal{C}_{\kappa}\mathcal{L}(x)\mathcal{C}_{\kappa}^\dagger &=& 
\mathcal{L}(x), \label{charge1}\\
\mathcal{C}_{\kappa}H\mathcal{C}_{\kappa}^\dagger &=& 
H, \label{charge2} \\
\mathcal{C}_{\kappa}\mathbf{P}\mathcal{C}_{\kappa}^\dagger &=& 
\mathbf{P}, \\
\mathcal{C}_{\kappa}\mathbf{J}\mathcal{C}_{\kappa}^\dagger &=& 
\mathbf{J}. \label{charge4}
\end{eqnarray}
These are the expected 
transformation laws for Eqs.~(\ref{adeld2})-(\ref{jvec2}) when we apply the charge 
conjugation operation. They are the same to the ones we have for the usual Dirac fields
\cite{gre95}. Note that to arrive at the above relations, we have discarded 
a four-divergence in Eq.~(\ref{charge1}) that does not change the action (\ref{acao}) 
and we have neglected spatial volume integrals
in Eqs.~(\ref{charge2})-(\ref{charge4}). These volume integrals 
can be transformed to surface integrals that 
are zero because the fields go to zero sufficiently
fast as we approach the surface of integration at infinity.

Let us now obtain the transformation law for the creation and annihilation operators
when subjected to the charge conjugation operation. 
Inserting Eq.~(\ref{psi}) into the left hand side
of Eq.~(\ref{chargeOp}) and making use of Eq.~(\ref{k2}) we get
\begin{eqnarray}
\mathcal{C}_{\kappa} \Psi(x)\mathcal{C}_{\kappa}^\dagger &=&
\sum_{r=1}^2\int \widetilde{dp}
[\mathcal{C}_{\kappa}c_r(\mathbf{p})\mathcal{C}_{\kappa}^\dagger]    
u_r(\mathbf{p})e^{-i\kappa x}e^{-ipx/\hbar} \nonumber \\
& + &\!\! \sum_{r=1}^2\int \widetilde{dp}
[\mathcal{C}_{\kappa}d^\dagger_r(\mathbf{p})\mathcal{C}_{\kappa}^\dagger] 
v_r(\mathbf{p})e^{-i\kappa x}e^{ipx/\hbar}. \nonumber \\
& & \label{lecharge}
\end{eqnarray}

On the other hand, inserting Eq.~(\ref{psi}) into the right hand side of
Eq.~(\ref{chargeOp}) we arrive at
\begin{eqnarray}
C\overline{\Psi}^T(x) \hspace{-.2cm}&=&\hspace{-.2cm}
\sum_{r=1}^2\int \widetilde{dp}
[ie^{i\varphi_{\!_C}}(-1)^{r}c_r^\dagger(\mathbf{p})] \nonumber \\    
&&\hspace{1cm}\times v_r(\mathbf{p})e^{-i\kappa x}e^{ipx/\hbar} \nonumber \\
\hspace{-.2cm}& + &\hspace{-.2cm} \sum_{r=1}^2\int \widetilde{dp}
[ie^{i\varphi_{\!_C}}(-1)^{r}d_r(\mathbf{p})] \nonumber \\ 
&&\hspace{1cm}\times u_r(\mathbf{p})e^{-i\kappa x}e^{-ipx/\hbar}. \nonumber \\
& & \label{ldcharge}
\end{eqnarray}
To obtain Eq.~(\ref{ldcharge}) we used the following identities, valid in the Dirac-Pauli
representation for the gamma matrices,
\begin{eqnarray}
\gamma^2u^*_r(\mathbf{p}) = (-1)^rv_r(\mathbf{p}), \\
\gamma^2v^*_r(\mathbf{p}) = (-1)^ru_r(\mathbf{p}).
\end{eqnarray}

Thus, comparing Eqs.~(\ref{lecharge}) and (\ref{ldcharge}) we get
\begin{eqnarray}
\mathcal{C}_{\kappa}c_r(\mathbf{p})\mathcal{C}_{\kappa}^\dagger &=&
ie^{i\varphi_{\!_C}}(-1)^{r}d_r(\mathbf{p}), \label{cc}\\
\mathcal{C}_{\kappa}d_r^\dagger(\mathbf{p})\mathcal{C}_{\kappa}^\dagger &=&
ie^{i\varphi_{\!_C}}(-1)^{r} c_r^\dagger(\mathbf{p}). \label{ddaggerc}
\end{eqnarray}
In Eqs.~(\ref{cc}) and (\ref{ddaggerc}) we have, similarly to Eqs.~(\ref{ct}) and (\ref{ddaggert}), transformation rules that depend on the spin index $r$. 
This is a feature of the asymmetric Dirac fields and does not happen for the usual
Dirac fields \cite{gre95}. 

We also have the following relations that are consequences of Eqs.~(\ref{cc}) and (\ref{ddaggerc}), 
\begin{eqnarray}
\mathcal{C}_{\kappa}c_r^\dagger(\mathbf{p})c_r(\mathbf{p})\mathcal{C}_{\kappa}^\dagger
&=&d_r^\dagger(\mathbf{p})d_r(\mathbf{p}), \\
\mathcal{C}_{\kappa}d_r^\dagger(\mathbf{p})d_r(\mathbf{p})\mathcal{C}_{\kappa}^\dagger
&=&c_r^\dagger(\mathbf{p})c_r(\mathbf{p}),
\end{eqnarray}
which, together with
\begin{equation}
\mathcal{C}_{\kappa} \overline{\Psi}(x)\mathcal{C}_{\kappa}^\dagger =
-\Psi^T(x)C^\dagger,
\label{chargeOpBar}
\end{equation}
are useful in proving Eqs.~(\ref{charge1})-(\ref{charge4}).

\subsection{The CPT operation}

If we successively apply the three discrete symmetry operations just studied, 
namely, Eqs.~(\ref{cp}), (\ref{ddaggerp}), (\ref{ct}), (\ref{ddaggert}), 
(\ref{cc}), and (\ref{ddaggerc}), and use that 
$(-1)^{\overline{r}}(-1)^{r+1}=1$, we have
\begin{eqnarray}
\Theta_{\kappa} c_r(\mathbf{p}) \Theta_{\kappa}^\dagger
&=& -e^{i(\varphi_{\!_C}+\varphi_{\!_P}+\varphi_{\!_T})}
d_{\overline{r}}(\mathbf{p}), \label{cptc}\\
\Theta_{\kappa} d^\dagger_r(\mathbf{p}) \Theta_{\kappa}^\dagger 
&=& e^{i(\varphi_{\!_C}+\varphi_{\!_P}+\varphi_{\!_T})}
c^\dagger_{\overline{r}}(\mathbf{p}), \label{cptd}
\end{eqnarray}
where
\begin{equation}
\Theta_{\kappa} = \mathcal{C}_{\kappa} \mathcal{P}_{\kappa} \mathcal{T}_{\kappa} 
\label{cptop}
\end{equation}
is the CPT operator. Note that the order we apply the operations $\mathcal{C}_{\kappa}, \mathcal{P}_{\kappa},$ and $\mathcal{T}_{\kappa}$ may affect the overall phase multiplying
$d_{\overline{r}}(\mathbf{p})$ and $c^\dagger_{\overline{r}}(\mathbf{p})$ in 
Eqs.~(\ref{cptc}) and (\ref{cptd}). However, the bilinears below transform in exactly the 
same way, irrespective of the order we apply those operations,
\begin{eqnarray}
\Theta_{\kappa} c^\dagger_r(\mathbf{p})c_r(\mathbf{p}) \Theta_{\kappa}^\dagger &=&
d^\dagger_{\overline{r}}(\mathbf{p})d_{\overline{r}}(\mathbf{p}),
\label{cptcc}\\
\Theta_{\kappa} d^\dagger_r(\mathbf{p})d_r(\mathbf{p}) \Theta_{\kappa}^\dagger &=&
c^\dagger_{\overline{r}}(\mathbf{p})c_{\overline{r}}(\mathbf{p}).
\label{cptdd}
\end{eqnarray}

Furthermore, using Eqs.~(\ref{parityOp}), (\ref{timereversalOp}),
and (\ref{chargeOp}) we obtain
\begin{eqnarray}
\Theta_{\kappa} \Psi(x) \Theta_{\kappa}^\dagger &=& 
e^{i(\varphi_{\!_C}+\varphi_{\!_P}+\varphi_{\!_T})}
i\gamma^0\gamma^5\overline{\Psi}^T(-x), \label{cptpsi}\\
\Theta_{\kappa} \overline{\Psi}(x) \Theta_{\kappa}^\dagger &=& 
e^{-i(\varphi_{\!_C}+\varphi_{\!_P}+\varphi_{\!_T})}
\Psi^T(-x)i\gamma^5\gamma^0. \label{cptpsibar}
\end{eqnarray}
If we set $\varphi_{\!_C}+\varphi_{\!_P}+\varphi_{\!_T}=0,\pm 2\pi, \pm 4\pi, \ldots$,
Eqs.~(\ref{cptpsi}) and (\ref{cptpsibar}) are formally equal to the 
way one usually writes how the standard Dirac fields transform 
under the CPT operation \cite{gre95}. Here we also have that the order in which we
apply the time reversal, parity, and charge conjugation operations may lead to a 
different phase at the right hand sides of Eqs.~(\ref{cptpsi}) and (\ref{cptpsibar}).
However, the transformation rules for the Dirac bilinears 
$\Psi(x)\Gamma \overline{\Psi}(x)$, where 
$\Gamma=\mathbb{1},\gamma^5,\gamma^\mu,\gamma^5\gamma^\mu,\sigma^{\mu\nu}$, 
are not affected by the order we apply those operations. 

Using Eqs.~(\ref{cptpsi}) and (\ref{cptpsibar}), or Eqs.~(\ref{cptc})-(\ref{cptd})
and (\ref{cptcc})-(\ref{cptdd}) if we work directly with the creation and annihilation
operators, we have
\begin{eqnarray}
\Theta_{\kappa}\mathcal{L}(x)\Theta_{\kappa}^\dagger &=& \mathcal{L}(-x), \label{cpt1}\\
\Theta_{\kappa}H\Theta_{\kappa}^\dagger &=& H, \\
\Theta_{\kappa}\mathbf{P}\Theta_{\kappa}^\dagger &=& \mathbf{P}, \\
\Theta_{\kappa}\mathbf{J}\Theta_{\kappa}^\dagger &=& -\mathbf{J}, \label{cpt4}
\end{eqnarray}
where the above quantities are given by Eqs.~(\ref{adeld2})-(\ref{jvec2}).
In writing Eq.~(\ref{cpt4}) as shown above, we used that the total angular momentum 
is conserved for a free field, i.e., $\mathbf{J}(-x^0)=\mathbf{J}(x^0)$. Also,
Eq.~(\ref{cpt1}) implies that the action (\ref{acao}) is invariant after 
the CPT transformation \cite{gre00,rig22,gre95},
\begin{equation}
\Theta_{\kappa}S\Theta_{\kappa}^\dagger = S.
\label{cpttheorem}
\end{equation}
This is an expected result since the CPT theorem guarantees that a local quantum
field theory canonically quantized and described by a Hermitian and Lorentz-invariant Lagrangian density should satisfy Eq.~(\ref{cpttheorem}). The asymmetric Dirac field
theory here presented satisfies all these assumptions.

If we use Eqs.~(\ref{pk1}), (\ref{tk1}), and (\ref{ck2}), together with the fact that
$(K_1)^2=\mathbb{1}$ and that $K_1$ commutes with $\mathcal{P}$, we can write 
Eq.~(\ref{cptop}) as follows,
\begin{equation}
\Theta_{\kappa} = K_2\mathcal{C} \mathcal{P} \mathcal{T} = K_2 \Theta, 
\label{cptm}
\end{equation}
where we can understand the operator $\Theta$ as the standard CPT operator associated
with the usual Dirac fields.

Remembering that [cf. Eq.~(\ref{k2})]
\begin{equation}
K_2 \kappa^\mu K_2^\dagger = - \kappa^\mu,
\end{equation}
we can understand Eq.~(\ref{cptm}) as a sort of
``CPTM'' operation. The ``M'' means that the parameters responsible to generate the 
mass of a particle, namely, $\kappa^\mu$, change sign after 
we implement the symmetry operation given by Eq.~(\ref{cptm}). In addition to the usual
$\Theta$ operator, in the theoretical framework
of the asymmetric Dirac fields we also have to include $K_2$ to arrive at a meaningful 
CPT operator that behaves as we expect and at a theory respecting the CPT theorem. 
As we already
pointed out several times, the inertial mass $m$, being a quadratic function
of $\kappa^\mu$, is not affected by the action of $K_2$. In other words, the mass $m$
does not change sign when subjected to the CPT operation as defined by Eq.~(\ref{cptm}); only $\kappa^\mu$ does change sign.\footnote{
This distinct behavior of $m$ and $\kappa^\mu$ under the CPT operation opens up at least
one interesting possibility deserving further investigation. Preliminary analysis show
that it might be possible to build a consistent gravitational quantum field theory
where particles and antiparticles repel each other gravitationally \cite{rig22}. This 
comes about by assuming that the gravitational field couples to $\kappa^0$
instead of $m$. In this scenario, particles attract particles, antiparticles attract 
antiparticles, and particles repel antiparticles gravitationally \cite{rig22} without the 
need to assume a negative mass for antiparticles \cite{saf18,bon57,kow96,vil11,far18}. Particles and antiparticles will always
have positive masses given by $m$ and positive energies, 
while particles will be associated with $\kappa^\mu$ and antiparticles with $-\kappa^\mu$.}

\subsection{The Dirac bilinears}

If we look at the right hand sides of Eqs.~(\ref{chargeOp}), (\ref{chargeOpBar}),
(\ref{cptpsi}) and (\ref{cptpsibar}), we realize that they are formally equal to the
way the standard Dirac fields transform under the charge conjugation and the CPT 
operations. Thus, since the Dirac bilinears do not depend on $\kappa^\mu$, the
transformation rules for them after those two operations
in the framework of the asymmetric Dirac theory are the same we obtain for the 
usual Dirac theory.

When it comes to the parity and time reversal operations, three of the five classes of 
Dirac bilinears transform differently in the framework of the asymmetric Dirac 
theory when compared to the way they transform in the context of the standard Dirac
theory. This different behavior can be traced back to the presence of the $\gamma^5$
matrix in the transformation rules for $\Psi(x)$ after the parity and time reversal
operations [cf. the right hand sides of Eqs.~(\ref{parityOp}) and (\ref{timereversalOp})], 
a feature that is absent in the transformation rules for standard Dirac
fields \cite{gre00,gre95}.

If we use Eqs.~(\ref{parityOp}), (\ref{timereversalOp}), and the respective expressions 
for the adjoint field,
\begin{eqnarray}
\mathcal{P}_{\kappa} \overline{\Psi}(x)\mathcal{P}_{\kappa}^\dagger &=&
-\overline{\Psi}(\widetilde{x})P^\dagger\gamma^5, \label{parityOpbar} \\
\mathcal{T}_{\kappa} \overline{\Psi}(x)\mathcal{T}_{\kappa}^\dagger &=&
-\overline{\Psi}(-\widetilde{x})T_0^\dagger\gamma^5,
\label{timereversalOpbar}
\end{eqnarray}
where $\widetilde{x}=(x^0,-\mathbf{r})$, we obtain that $\overline{\Psi}(x)\Psi(x)$
behaves as a pseudoscalar and $\overline{\Psi}(x)i\gamma^5\Psi(x)$ as a scalar under 
the parity and time reversal operations. Also, 
$\overline{\Psi}(x)\sigma^{\mu\nu}\Psi(x)$ behaves as a pseudotensor when subjected to
those two transformations. This should be contrasted to the behavior of these 
three class of bilinears in the framework of the standard Dirac theory, where they
behave, respectively, as a scalar, pseudoscalar, and a tensor. In the present 
theoretical framework, a tensor under those two transformations is given by 
$\overline{\Psi}(x)i\gamma^5\sigma^{\mu\nu}\Psi(x)$.
On the other hand,
similar calculations show that $\overline{\Psi}(x)\gamma^\mu\Psi(x)$ and 
$\overline{\Psi}(x)\gamma^5\gamma^\mu\Psi(x)$ transform under those symmetry operations
in exactly the same way the corresponding quantities do in the framework of the standard
Dirac theory. We have that $\overline{\Psi}(x)\gamma^\mu\Psi(x)$ behaves as a 
vector and that $\overline{\Psi}(x)\gamma^5\gamma^\mu\Psi(x)$ behaves 
as a pseudovector when subjected to those symmetry operations.

We should also mention that $\kappa^0, \kappa^1, \kappa^2, \kappa^3$ are 
four relativistic invariants only under proper Lorentz transformations. Under the discrete 
symmetry operations here investigated, and combinations thereof, $\kappa^\mu$ behaves
as a regular four-vector, changing sign when subjected to those symmetry operations according to the following rule \cite{rig23},
\begin{eqnarray}
\kappa^\mu &\xrightarrow{\mbox{\tiny{parity}}}& \kappa_\mu, \label{parityF}\\
\kappa^\mu &\xrightarrow{\mbox{\tiny{time reversal}}}& \kappa_\mu, \label{trF}\\
\kappa^\mu &\xrightarrow{\mbox{\tiny{charge conj.}}}& -\kappa^\mu. \label{cgF}
\end{eqnarray}

Finally, since the transformation rules for the asymmetric Dirac fields under the CPT 
operation are formally equal to the ones related to the standard Dirac fields
[cf. Eqs.~(\ref{cptpsi}) and (\ref{cptpsibar})], 
the analysis leading to the equality of masses and lifetimes for particles and antiparticles
\cite{gre95} can be carried over to the present theoretical framework 
if we identify the mass of particles and antiparticles as given by Eq.~(\ref{restmass}), 
a genuine invariant under proper and improper Lorentz transformations. These results,
together with the analysis made in Secs.~\ref{fp} and \ref{qed}, 
lead to the proof of the empirical equivalence between 
any Lorentz invariant abelian gauge theory 
built within the theoretical framework of the asymmetric Dirac fields and 
within the theoretical framework of the usual Dirac fields.

\section{The Feynman propagator}
\label{fp}

\subsection{Arbitrary time anticommutator}

To obtain the Feynman propagator in the canonical formalism, we first need to compute
the following anticommutator for arbitrary space-time points,
\begin{eqnarray}
\{\Psi(x),\overline{\Psi}(y)\} &=& \{\Psi^+(x)+\Psi^-(x),
\overline{\Psi}^+(y)+\overline{\Psi}^-(y)\} \nonumber \\
&=& \{\Psi^+(x),\overline{\Psi}^-(y)\} + \{\Psi^-(x),\overline{\Psi}^+(y)\}. \nonumber \\
&& \label{antiarbitrary}
\end{eqnarray}
We should think of Eq.~(\ref{antiarbitrary}) as a $4\times 4$ matrix equation in the 
spinor space. For simplicity, we have omitted the spin indexes. And to arrive at its 
last line, we used that $\{\Psi^+(x),\overline{\Psi}^+(y)\}=
\{\Psi^-(x),\overline{\Psi}^-(y)\}=0$, which can easily be proved using 
Eqs.~(\ref{psi+})-(\ref{psibar-}) and that 
$\{c_r(\mathbf{p}),d_s(\mathbf{p'})\}
=\{d^\dagger_r(\mathbf{p}),c^\dagger_s(\mathbf{p'})\}=0$.

The remaining non-trivial anticommutators are computed using Eqs.~(\ref{psi+})-(\ref{psibar-}) and (\ref{aad}). A direct calculation gives
\begin{eqnarray}
\{\Psi^+(x),\overline{\Psi}^-(y)\} &=& e^{i\kappa (x-y)}\int d^3p g^2(|\mathbf{p}|)
e^{-ip(x-y)/\hbar} \nonumber \\
&&\times \left(\sum_{r=1}^2 u_r(\mathbf{p})\overline{u}_r(\mathbf{p})\right), 
\label{psi+-}\\
\{\Psi^-(x),\overline{\Psi}^+(y)\} &=& e^{i\kappa (x-y)}\int d^3p g^2(|\mathbf{p}|)
e^{ip(x-y)/\hbar} \nonumber \\
&&\times \left(\sum_{r=1}^2 v_r(\mathbf{p})\overline{v}_r(\mathbf{p})\right)\label{psi-+}.
\end{eqnarray}

In Ref.~\cite{rig23} we proved that
\begin{eqnarray}
\sum_{r=1}^2 u_r(\mathbf{p})\overline{u}_r(\mathbf{p}) & = &
-i\Lambda^+(\mathbf{p})\gamma^5, \label{uubar}\\ 
\sum_{r=1}^2 v_r(\mathbf{p})\overline{v}_r(\mathbf{p}) & = &
i\Lambda^-(\mathbf{p})\gamma^5, \label{vvbar}
\end{eqnarray}
where the energy projection operators are given by \cite{rig23}
\begin{equation}
\Lambda^{\pm}(\mathbf{p})=\frac{(\pm \slashed p - imc \gamma^5)i\gamma^5}{2mc}. 
\label{lpm}
\end{equation}

Thus, inserting Eqs.~(\ref{uubar})-(\ref{lpm}) into Eqs.~(\ref{psi+-})-(\ref{psi-+}) and
using Eq.~(\ref{gp}) we get
\begin{equation}
\{\Psi^{\pm}(x),\overline{\Psi}^{\mp}(y)\} = i e^{i\kappa (x-y)} 
(i\slashed \partial_x - i \widetilde{m}\gamma^5)\Delta^{\pm}(x-y).
\label{psipm}
\end{equation}
Here $\widetilde{m}=mc/\hbar$ and $\Delta^{\pm}(x)$ are the usual 
$\Delta$-functions related to the commutators of the Klein-Gordon fields
\cite{rig22,man86,gre95},
\begin{equation}
\Delta^{\pm}(x) = -\Delta^{\mp}(-x) = \frac{\mp ic}{2(2\pi)^3}\int d^3k\, 
\frac{e^{\mp ikx}}{\omega_{\mathbf{k}}}, \label{deltapm2}
\end{equation}
with
\begin{eqnarray}
E_{\mathbf{p}} = c p^0 = \hbar \omega_{\mathbf{k}} = c\hbar k^0.
\label{wk}
\end{eqnarray}

We should remark that to arrive at Eq.~(\ref{psipm}) we used that
\begin{equation}
\slashed p e^{\pm i p(x-y)/\hbar} = 
\mp i \hbar\slashed \partial_x e^{\pm i p(x-y)/\hbar},
\label{delx}
\end{equation}
where the subscript $x$ in $\slashed \partial_x$ reminds us that we should 
differentiate with respect to $x$, i.e.,  
\begin{equation}
\slashed \partial_x=\gamma^\mu\frac{\partial}{\partial x^\mu}.
\end{equation}

If we now use Eq.~(\ref{psipm}), we can write Eq.~(\ref{antiarbitrary}) as follows,
\begin{equation}
\{\Psi(x),\overline{\Psi}(y)\}  =   i e^{i\kappa (x-y)} 
(i\slashed \partial_x - i \widetilde{m}\gamma^5)\Delta(x-y),
\label{antiarbitrary2}
\end{equation}
where
\begin{equation}
\Delta(x) = \Delta^{+}(x) + \Delta^{-}(x).  
\end{equation}

We can also cast Eq.~(\ref{antiarbitrary}) similarly to the way one writes the usual
Dirac fields anticommutators \cite{man86,gre95} if we define
\begin{eqnarray}
S^{\pm}(x) &=&  e^{i\kappa x} 
(i\slashed \partial - i \widetilde{m}\gamma^5)\Delta^\pm(x), \label{spm}\\
S(x) &=&  e^{i\kappa x}  
(i\slashed \partial - i \widetilde{m}\gamma^5)\Delta(x). \label{s}
\end{eqnarray}
Note that we are writing $\slashed \partial$ instead of $\slashed \partial_x$ since
now we only have the variable $x$ onto which the operator $\slashed \partial$ acts. 
Using Eqs.~(\ref{spm}) and (\ref{s}), Eq.~(\ref{antiarbitrary2}) becomes
\begin{equation}
\{\Psi(x),\overline{\Psi}(y)\}  =   i S^+(x-y)+ i S^-(x-y) = i S(x-y).  
\label{antiarbitrary3}
\end{equation}

We can also write $S^\pm(x)$ as a contour integral in the complex $p^0$-plane if we use 
that \cite{man86,gre95}
\begin{equation}
\Delta^\pm(x) = -\frac{\hbar^2}{(2\pi\hbar)^4}\int_{C^{\pm}}d^4p\,
\frac{e^{-ipx/\hbar}}{p^2-m^2c^2}, \label{contourCpm}
\end{equation}
where the complex contour integral in the $p_0$-plane is such that
$C^+$ (or $C^-$) is any counterclockwise closed
path encircling only $E_{\mathbf{p}}/c$ (or $-E_{\mathbf{p}}/c$), 
where $p^0=\pm E_{\mathbf{p}}/c$ are the two simple poles of the integrand. 

If we use Eqs.~(\ref{delx}) and (\ref{contourCpm}), we can write Eq.~(\ref{spm}) as
\begin{equation}
S^\pm(x) = -\frac{\hbar}{(2\pi\hbar)^4}\int_{C^{\pm}}d^4p\,
\frac{\slashed p -imc\gamma^5}{p^2-m^2c^2}e^{-ipx/\hbar}e^{i\kappa x}. 
\label{Spm}
\end{equation}
And since 
\begin{equation}
(\slashed p \pm imc\gamma^5)(\slashed p \pm imc\gamma^5)=p^2-m^2c^2,
\end{equation}
we can formally write
\begin{equation}
\frac{\slashed p -imc\gamma^5}{p^2-m^2c^2} =
\frac{\slashed p -imc\gamma^5}{(\slashed p -imc\gamma^5)(\slashed p -imc\gamma^5)} \equiv
\frac{1}{\slashed p -imc\gamma^5}. \label{symbolic}
\end{equation}

Therefore, Eq.~(\ref{Spm}) becomes
\begin{equation}
S^\pm(x) = -\frac{\hbar}{(2\pi\hbar)^4}\int_{C^{\pm}}d^4p\,
\frac{e^{-i(p-\hbar \kappa)x/\hbar}}{\slashed p -imc\gamma^5}. 
\label{Spm2}
\end{equation}

We should note that if in Eq.~(\ref{Spm2}) we choose any counterclockwise path that 
encircles both simple poles ($\pm E_{\mathbf{p}}/c$), we obtain the complex integral
representation for $S(x)$ as given by Eq.~(\ref{s}). Also, if we set $\kappa=0$ and 
change $imc\gamma^5$ to $mc$ in Eq.~(\ref{Spm2}), we obtain the analog 
expression for the standard Dirac fields.

\subsection{The fermion propagator}

The Feynman propagator for the asymmetric Dirac fields is defined in exactly 
the same way we define a propagator for any pair of fermion fields \cite{man86,gre95},
\begin{equation}
\langle 0| T\{\Psi_\alpha(x)\overline{\Psi}_\beta(x')\} 
\!|0\rangle = iS_{F_{\alpha\beta}}(x-x'), \label{FP}
\end{equation}
with $|0\rangle$ being the vacuum state and $\alpha,\beta$ spinor indexes. 

Suppressing from now on the spinor indexes, 
the time ordering operation is given by \cite{man86,gre95},
\begin{eqnarray}
T\{\Psi(x)\overline{\Psi}(x')\} &=& h(x^0-x^{0'})\Psi(x)\overline{\Psi}(x')
\nonumber \\
& & - 
h(x^{0'}-x^0)\overline{\Psi}(x')\Psi(x), \label{Tproduct}
\end{eqnarray}
where the Heaviside step function is,
\begin{eqnarray}
h(x^0) &=& 1, \;\mbox{if}\; x^0>0, \nonumber \\
h(x^0) &=& 0, \;\mbox{if}\; x^0<0.
\label{hfunction}
\end{eqnarray}

And since \cite{man86,gre95}
\begin{eqnarray}
\langle 0|\Psi(x)\overline{\Psi}(x')|0\rangle &=& i S^+(x-x'), 
\label{ppd}\\
\langle 0|\overline{\Psi}(x')\Psi(x)|0\rangle &=& i S^-(x-x'),
\label{pdp}
\end{eqnarray}
Eq.~(\ref{FP}) becomes
\begin{equation}
\langle 0| T\{\Psi_\alpha(x)\overline{\Psi}_\beta(x')\} \!|0\rangle = 
iS_F(x-x'), \label{FP2}
\end{equation}
where
\begin{equation}
S_F(x)=h(x^0)S^+(x)- h(-x^0)S^-(x). \label{sf}
\end{equation}

Inserting Eq.~(\ref{spm}) into (\ref{sf}) we obtain
\begin{equation}
S_F(x)=e^{i\kappa x}  
(i\slashed \partial - i \widetilde{m}\gamma^5)\Delta_F(x),
\label{sf2}
\end{equation}
where $\Delta_F(x)=h(x^0)\Delta^+(x)-h(-x^0)\Delta^-(x)$ is the 
propagator related to the Klein-Gordon complex scalar fields \cite{man86,gre95}.
Moreover, if we use the following identity,
\begin{equation}
e^{i\kappa x} i\slashed \partial \Delta_F(x) = \slashed \kappa e^{i\kappa x} \Delta_F(x)
+ i \slashed \partial [ e^{i\kappa x} \Delta_F(x)],
\end{equation}
Eq.~(\ref{sf2}) becomes
\begin{eqnarray}
S_F(x)&=&(i\slashed \partial - i \widetilde{m}\gamma^5 + \slashed \kappa)
[e^{i\kappa x}\Delta_F(x)], \nonumber \\
&=&(i\slashed \partial - i \widetilde{m}\gamma^5 + \slashed \kappa)
[\Delta_{F_{GLS}}(x)].
\label{sf3}
\end{eqnarray}
Here $\Delta_{F_{GLS}}(x)$ is the Feynman propagator of the generalized Lorentz covariant
Schr\"odinger fields \cite{rig22}. Note also that in Eq.~(\ref{sf3}) the operator
$(i\slashed \partial - i \widetilde{m}\gamma^5 + \slashed \kappa)$ is, up to an
overall $\hbar$, the operator
associated with the asymmetric Dirac equation [cf. Eq.~(\ref{adeFinal2})].

If in Eq.~(\ref{Spm2}) we slightly displace off the real $p^0$-axis 
the simple poles of the integrand, 
\begin{equation}
\pm \frac{E_\mathbf{p}}{c} \longrightarrow \pm \frac{E_\mathbf{p}}{c} \mp i\eta, 
\end{equation}
we can express the Feynman propagator such that 
all integrated variables are real numbers \cite{man86,gre95}, 
\begin{equation}
S_{F}(x) =\frac{\hbar}{(2\pi\hbar)^4}\int d^4p\,
\frac{\slashed p - imc\gamma^5}{p^2-m^2c^2+i\epsilon}e^{-i(p-\hbar\kappa)x/\hbar},
\label{sf4}
\end{equation}
where 
\begin{equation}
\epsilon = \frac{2\eta}{c}\frac{E_{\mathbf{p}}}{\hbar},
\end{equation}
with $0<\eta \ll 1$. In Eq.~(\ref{sf4}), $p^0$ is also integrated from 
$-\infty$ to $\infty$ and the limit $\eta\rightarrow 0$ (or equivalently 
$\epsilon \rightarrow 0$) is taken after the integration. 

Using the symbolic representation (\ref{symbolic}) for the kernel of the four dimensional Fourier transform given by Eq.~(\ref{sf4}), we obtain
\begin{equation}
S_{F}(x) =e^{i\kappa x}\frac{1}{(2\pi\hbar)^4}\int d^4p\,
S_F(p)e^{-ipx/\hbar},
\label{sf5}
\end{equation}
where
\begin{equation}
S_F(p)=\frac{\hbar}{\slashed p - imc\gamma^5+i\epsilon}.
\label{sfp}
\end{equation}

We should also mention that by employing Eq.~(\ref{sf4}), a direct calculation gives
\begin{equation}
(i\slashed \partial  - i\widetilde{m}\gamma^5 + \slashed \kappa )S_F(x) 
=e^{i\kappa x}\delta^{(4)}(x)= \delta^{(4)}(x), \label{green}
\end{equation}
where
\begin{equation}
 \delta^{(4)}(x) = \frac{1}{(2\pi\hbar)^4}\int d^4p\, e^{-ipx/\hbar}.
\end{equation}
Equation (\ref{green}) shows that, as expected, 
$S_F(x)$ is proportional to the Green's function of the 
asymmetric Dirac equation.

\subsection{An auxiliary propagator}

If we write the asymmetric Dirac field as
\begin{equation}
\Psi(x) = e^{i\kappa x} \widetilde{\Psi}(x),
\end{equation}
the asymmetric Dirac equation becomes \cite{rig23}
\begin{equation}
(i\hbar \slashed \partial -imc\gamma^5) \widetilde{\Psi}(x) = 0.
\label{tildeADE}
\end{equation}
Equation (\ref{tildeADE}) is formally equivalent to (\ref{adeFinal2}) if $\kappa = 0$. 
Thus, its propagator is readily obtained from that associated with the asymmetric Dirac
equation,
\begin{equation}
\langle 0| T\{\widetilde{\Psi}(x)\overline{\widetilde{\Psi}}(x')\} 
\!|0\rangle = i\widetilde{S}_{F}(x-x'), \label{FPtilde}
\end{equation}
where 
\begin{equation}
\widetilde{S}_{F}(x) =\frac{1}{(2\pi\hbar)^4}\int d^4p\,
\widetilde{S}_F(p)e^{-ipx/\hbar}
\label{sf5tilde}
\end{equation}
and
\begin{equation}
\widetilde{S}_F(p)=S_F(p)=\frac{\hbar}{\slashed p - imc\gamma^5+i\epsilon}.
\label{sfptilde}
\end{equation}

The Feynman propagator in momentum space $i\widetilde{S}_F(p)$ will be very useful
in formulating the Feynman rules of QED for the present theory in Sec.~\ref{qed}.

\subsection{Connection with the standard fermion propagator}

If we use Eq.~(\ref{aDtoD}) and its adjoint,
\begin{equation}
\overline{\Psi}(x)= e^{-i\kappa x} \overline{\Psi}_D(x)U,
\label{aDtoDadj}
\end{equation}
it is not difficult to see that
\begin{eqnarray}
\langle 0| T\{\Psi(x)\overline{\Psi}(x')\} 
\!|0\rangle = e^{i\kappa(x-x')}\nonumber \\ 
\times U\langle 0| T\{\Psi_D(x)\overline{\Psi}_D(x')\} 
\!|0\rangle U. \label{FPDirac}
\end{eqnarray}
Now, if we use Eq.~(\ref{FP}), and its analog related to the standard Dirac fields, we get
from Eq.~(\ref{FPDirac}) that
\begin{equation}
S_F(x)= e^{i\kappa x}U S_{D_F}(x)U, \label{FPDirac2}
\end{equation}
where 
\begin{equation}
S_{D_F}(x) =\frac{\hbar}{(2\pi\hbar)^4}\int d^4p\,
\frac{\slashed p +mc}{p^2-m^2c^2+i\epsilon}e^{-ipx/\hbar}
\label{sf4Dirac}
\end{equation}
is proportional to $iS_{D_F}(x)$, the propagator for the standard Dirac fields.

We can also directly check the validity of Eq.~(\ref{FPDirac2}) using the explicit forms
for $S_F(x)$ and $S_{D_F}(x)$, given by Eqs.~(\ref{sf4}) and (\ref{sf4Dirac}), and
the following identities,
\begin{eqnarray}
U \gamma^\mu U =\gamma^\mu U^\dagger U= \gamma^\mu, & U \gamma^5 U = \gamma^5 U^2 = -i.  
\label{uidentities}
\end{eqnarray}

And if we note that $S_F(x)=e^{i\kappa x}\widetilde{S}_{F}(x)$, 
Eq.~(\ref{FPDirac2}) leads to
\begin{equation}
\widetilde{S}_{F}(x)= U S_{D_F}(x)U. \label{FPDiracTilde}
\end{equation}

\section{Quantum electrodynamics}
\label{qed}

\subsection{The interaction Hamiltonian density}
\label{equalham}

The kinetic term of the Lagrangian density describing the asymmetric Dirac fields 
is formally equal to the kinetic term of the standard Dirac Lagrangian density [cf. Eq.~(\ref{adeld2})].
Therefore, if we apply the minimal coupling prescription to Eq.~(\ref{adeld2}) we will
obtain the same interaction term associated with the standard QED. 

The minimal coupling prescription \cite{gre00,man86,gre95} in SI units and in the present
metric signature is \cite{rig22,rig23}
\begin{equation}
\partial_\mu \rightarrow D_\mu = \partial_\mu + \frac{iq}{\hbar}A_\mu(x),
\label{mc}
\end{equation}
where $q$ is the electric charge of the particle ($q=-e<0$ for an electron) and the covariant four-vector potential is 
\begin{equation}
A_\mu = \left( \frac{\varphi}{c}, -\mathbf{A} \right). 
\label{pres}
\end{equation}
Here $\varphi$ and $\mathbf{A}=(A^1,A^2,A^3)$ correspond to, respectively,
the electric and vector potentials associated with an electromagnetic field.
In the second quantization formalism, $A^\mu$ becomes an operator expanded in terms of 
photon creation and annihilation operators \cite{man86,gre95}.

Applying the minimal coupling prescription (\ref{mc}) to the normal ordered version of Eq.~(\ref{adeld2}), we obtain the following interaction term,
\begin{equation}
\mathcal{L}_{int}(x) = -cq :\overline{\Psi}(x)\slashed A(x) \Psi(x):.
\label{linta}
\end{equation}
As anticipated above, $\mathcal{L}_{int}(x)$ 
is formally equal to the interaction Lagrangian density
of the usual Dirac fields. Moreover, if we use the expressions connecting the asymmetric
Dirac fields with the standard ones, Eqs.~(\ref{aDtoD}) and (\ref{aDtoDadj}), and the identity given by Eq.~(\ref{uidentities}), we can write Eq.~(\ref{linta}) as follows,
\begin{equation}
\mathcal{L}_{int}(x) = -cq :\overline{\Psi}_D(x)\slashed A(x) \Psi_D(x): 
= \mathcal{L}_{D_{int}}(x),
\label{lintd}
\end{equation}
where $\mathcal{L}_{D_{int}}(x)$ is the interaction term of the standard QED. 
Equation (\ref{lintd}) is one of the key results that allow us to prove the equivalence
between the standard QED and the asymmetric Dirac field QED.

Since we will be working in the interaction picture, from now on we assume the 
fields are already expressed in that picture. And since
Eq.~(\ref{linta}) does not depend on time derivatives, in the interaction picture 
the Hamiltonian density modeling the electromagnetic interaction is given by 
\begin{equation}
\mathcal{H}_{int}(x)  = - \mathcal{L}_{int}(x) = cq :\overline{\Psi}(x)\slashed A(x) \Psi(x):.
\label{hinta}
\end{equation}

We should also mention that the asymmetric Dirac fields respect lepton universality.  
If we include all leptons in the free field Lagrangian, the minimal coupling prescription
leads to the usual interaction Hamiltonian density for electrically charged leptons \cite{man86,gre95},
\begin{equation}
\mathcal{H}_{L_{int}}(x)  = - \mathcal{L}_{L_{int}}(x) = 
cq \sum_{l}:\overline{\Psi_l}(x)\slashed A(x) \Psi_l(x):,
\label{hlinta}
\end{equation}
where $l$ labels the leptons involved in the process investigated.

Before we move on, it is important at this moment to highlight that the equivalence
between the standard and the asymmetric Dirac field interaction Lagrangian densities, 
as given by Eq.~(\ref{lintd}), or between the respective interaction Hamiltonian densities, is a consequence of two facts. The first one is the particular form of Eq.~(\ref{aDtoD}), the time-dependent unitary transformation formally connecting the two theories. The second one is the specific form of the Hamiltonian density describing
the electromagnetic interaction. Indeed, since the latter does not have any 
space-time derivatives, the phase $e^{i\kappa x}$ related to 
$\Psi(x)$ is straightforwardly canceled by the phase $e^{-i\kappa x}$ coming from 
$\overline{\Psi}(x)$ when we insert Eqs.~(\ref{aDtoD}) and (\ref{aDtoDadj})
into (\ref{linta}). 
Also, since the electromagnetic interaction is proportional to
$\gamma^\mu$ and since $U\gamma^\mu U = \gamma^\mu$, no change occurs in the 
electromagnetic interaction Hamiltonian density 
when we use Eqs.~(\ref{aDtoD}) and (\ref{aDtoDadj}) to go
from the asymmetric to the standard Dirac fields. For interaction Hamiltonian
densities involving space-time
derivatives of the fields or bilinears $\Gamma$ that do not satisfy
$U\Gamma U = \Gamma$, we will have a different interaction Hamiltonian density when we
employ Eqs.~(\ref{aDtoD}) and (\ref{aDtoDadj}).

\subsection{The S-matrix, Feynman amplitudes, and the cross section}
\label{smatrix}

Even though the interaction Hamiltonian densities of the asymmetric Dirac field QED and
the standard QED were shown to be same in Sec. \ref{equalham}, we have to check
whether they lead to the same experimental predictions within the theoretical framework
of those two theories. If we go through the perturbative techniques used to 
compute cross sections, we realize that the same steps leading to the Dyson series and
the application of Wick's theorem in the context of the standard QED are the same 
for the asymmetric Dirac field QED.

Furthermore, we can make any Feynman amplitude, transition matrix, and cross section equal in both theories if 
\begin{enumerate}
\item[(A)] We assume the map given by Eq.~(\ref{mapspinors}) 
between the standard and asymmetric Dirac
spinors.
\item[(B)] We identify a standard Dirac particle (antiparticle) 
with four-momentum $p^\mu$ $(p^\mu)$ with an asymmetric Dirac particle (antiparticle)
with four-wave number  $k^\mu=p^\mu/\hbar$ whose energy is 
$E_{\mathbf{p}} - c\hbar \kappa^0$ $(E_{\mathbf{p}} + c\hbar \kappa^0)$ and whose momentum is $\mathbf{p} - \hbar \bm{\kappa}$ $(\mathbf{p} + \hbar \bm{\kappa})$. 
\item[(C)] We identify the velocity of an asymmetric Dirac particle (antiparticle)
with four-wave number  $k^\mu=p^\mu/\hbar$ as given
by $\mathbf{p}/m$ $(\mathbf{p}/m)$.
\end{enumerate}

The map required by point (A) above is 
\begin{eqnarray}
w_r(\mathbf{p}) = U w_{\!_Dr}(\mathbf{p}), & 
\overline{w}_r(\mathbf{p}) = \overline{w}_{\!_Dr}(\mathbf{p}) U, 
\label{mapspinors}
\end{eqnarray}
where $U$ is given by Eq.~(\ref{U}), $w=u,v$, and the subscript $D$ reminds us that we are dealing with the standard Dirac spinors. Equation (\ref{mapspinors}) is the analog in
momentum space of Eq.~(\ref{aDtoD}), the map in position space connecting 
$\Psi(x)$ with $\Psi_{D}(x)$. 

This map allows us to go from the asymmetric Dirac equation in momentum space to the 
standard Dirac equation in momentum space. For instance, if we insert Eq.~(\ref{mapspinors})
into (\ref{ue}), we get after left multiplying by $U$ and using Eq.~(\ref{uidentities}),
\begin{equation}
(\slashed p -mc) u_{\!_Dr}(\mathbf{p}) = 0,
\end{equation}
which is exactly the standard Dirac equation in momentum space for 
$u_{\!_Dr}(\mathbf{p})$. Repeating the previous calculation using Eq.~(\ref{ve}), 
we obtain $(\slashed p +mc) v_{\!_Dr}(\mathbf{p}) = 0$,
the standard Dirac equation in momentum space for 
$v_{\!_Dr}(\mathbf{p})$. 

The map given by Eq.~(\ref{mapspinors}) ensures that a Feynman amplitude for a given 
process computed within the theoretical framework of the asymmetric Dirac fields 
is exactly mapped to the corresponding Feynman amplitude obtained within the theoretical framework of the standard Dirac fields. Point (B) above guarantees that the S-matrix
elements representing the probability amplitude for a given process are properly 
mapped from one theory to the other one. Finally, point (C) is justified when the   
incident flux for a given scattering process is computed in both theories. The equality
of those fluxes leads to point (C) and ultimately to the equality in both theories of the 
cross section for a given process.

\subsubsection*{Compton scattering}

\begin{figure}[!ht]
\centering 
\includegraphics[width=8.5cm]{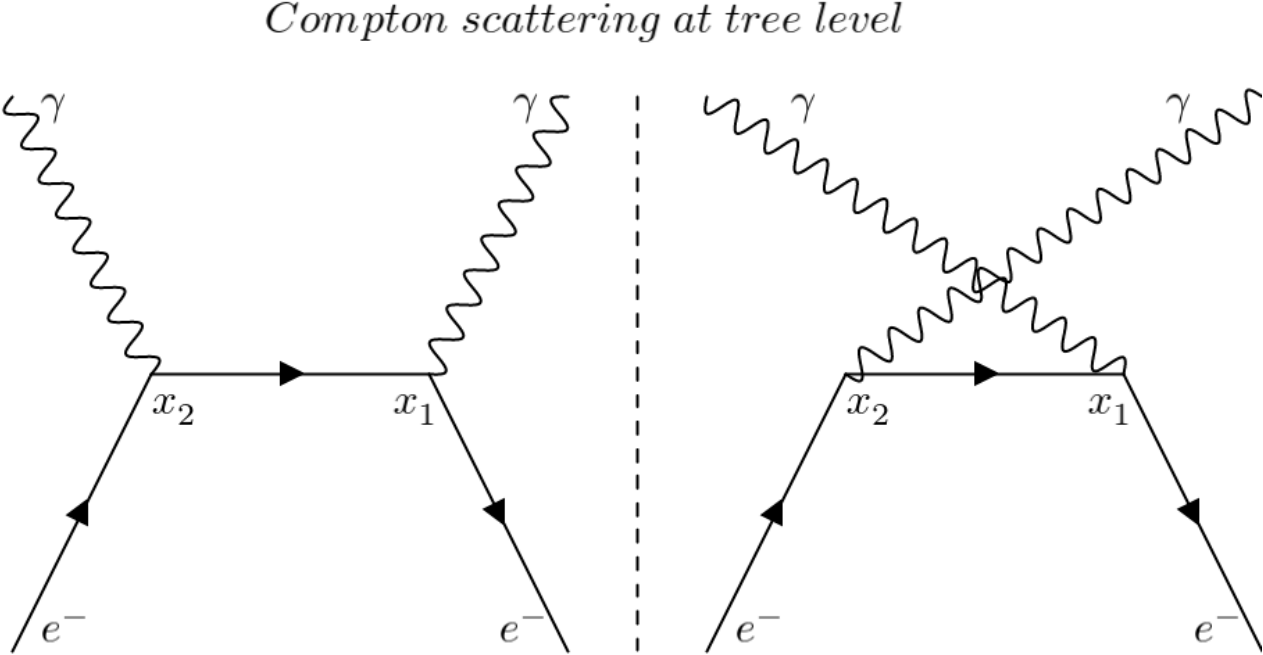}
\caption{\label{fig1} Feynman diagrams in position
space related to Compton scattering 
at the lowest order in perturbation theory.}
\end{figure}

To illustrate the points just raised in Sec. \ref{smatrix}, 
we compute the transition matrix for a particular scattering process.
Specifically, we will now be dealing with the computation of the transition matrix
elements for Compton scattering at tree level. The 
relevant Feynman diagrams in position space are given by Fig.~\ref{fig1}.

If we use the same notation and repeat for the present theory 
the same mathematical steps of Ref. \cite{man86} when it dealt with 
Compton scattering, we obtain that the part of the 
perturbative expansion of the S-matrix associated with the Feynman diagrams of 
Fig.~\ref{fig1} is
\begin{equation}
S^{(2)}(\gamma e^- \rightarrow \gamma e^-) = S_a + S_b,
\end{equation}
where in SI units we have
\begin{eqnarray}
S_a &=& 2\left(\frac{-i}{\hbar c}\right)^2\frac{1}{2!}(-ec)^2\int d^4x_1 d^4x_2 
\overline{\Psi}^-(x_1)\gamma^\mu  \nonumber \\
&\times& iS_F(x_1-x_2)\gamma^\nu A^-_\mu(x_1)
A^+_\nu(x_2)\Psi^+(x_2), \label{Sa}\\
S_b  &=& 2\left(\frac{-i}{\hbar c}\right)^2\frac{1}{2!}(-ec)^2\int d^4x_1 d^4x_2 
\overline{\Psi}^-(x_1)\gamma^\mu \nonumber \\
&\times&  iS_F(x_1-x_2)\gamma^\nu A^-_\nu(x_2)
A^+_\mu(x_1)\Psi^+(x_2). \label{Sb}
\end{eqnarray}
In Eqs.~(\ref{Sa}) and (\ref{Sb}), $\Psi^\pm(x_j)$ and $\overline{\Psi}^\pm(x_j)$, 
where
$j=1,2$, are given
by Eqs.~(\ref{psi+})-(\ref{psibar-}), $A^\pm_\mu(x_j)$ are the corresponding photon field
expansions \cite{man86}, and $iS_F(x_1-x_2)$ is the fermion 
propagator given by Eq.~(\ref{sf5}).

If we look at Eqs.~(\ref{psi+})-(\ref{psibar-}) and (\ref{sf5}), we realize that
the exponentials $e^{i\kappa x}$ coming from $\Psi^\pm(x)$, the exponentials
$e^{-i\kappa x}$ coming from $\overline{\Psi}^\pm(x)$, and the exponential
$e^{i\kappa x}$ of $S_F(x)$ are all canceled in Eqs.~(\ref{Sa}) and (\ref{Sb}).
As such, the mathematical steps leading to the transition matrix for Compton scattering
are formally the same of standard QED. Therefore, following \cite{man86} we get
\begin{eqnarray}
\langle f | S_a | i \rangle &=&  (2\pi\hbar)^4\delta^{(4)}(p'+\hbar k' - p - \hbar k)
\nonumber \\
&&\times g(|\mathbf{p'}|)g(|\mathbf{p}|)f(|\mathbf{k'}|)f(|\mathbf{k}|)\mathcal{M}_a,
\label{Sa2}\\
\langle f | S_b | i \rangle &=&  (2\pi\hbar)^4\delta^{(4)}(p'+\hbar k' - p - \hbar k)
\nonumber \\
&&\times g(|\mathbf{p'}|)g(|\mathbf{p}|)f(|\mathbf{k'}|)f(|\mathbf{k}|)\mathcal{M}_b,
\label{Sb2}
\end{eqnarray}
where, in an obvious notation and suppressing the spin and polarization indexes, 
$|i\rangle=c^\dagger(\mathbf{p})a^\dagger(\mathbf{k})|0\rangle, |f\rangle=
c^\dagger(\mathbf{p'})a^\dagger(\mathbf{k'})|0\rangle$, and
\begin{eqnarray}
\mathcal{M}_a &=& -\frac{e^2}{\hbar^2}\overline{u}(\mathbf{p'})\slashed \varepsilon(\mathbf{k'})
i\widetilde{S}_F(p+\hbar k)
\slashed \varepsilon(\mathbf{k})u(\mathbf{p}),
\label{ma}\\
\mathcal{M}_b &=& -\frac{e^2}{\hbar^2}\overline{u}(\mathbf{p'})\slashed \varepsilon(\mathbf{k})
i\widetilde{S}_F(p-\hbar k')
\slashed \varepsilon(\mathbf{k'})u(\mathbf{p}).
\label{mb}
\end{eqnarray}

In Eqs.~(\ref{Sa2}) and (\ref{Sb2}), $g(|\mathbf{p}|)$ is given by Eq.~(\ref{gp}) and
$f(|\mathbf{k}|)$ is the corresponding normalization factor for the photon field 
expansion. In SI units, 
\begin{equation}
f(|\mathbf{k}|)=\sqrt{\frac{\mu_0\hbar c^2}{2(2\pi)^3\omega_{\mathbf{k}}}}, 
\end{equation}
where $\mu_0$ is the vacuum permeability and $\omega_{\mathbf{k}}$ is defined in 
Eq.~(\ref{wk}).

In Eqs.~(\ref{ma}) and (\ref{mb}), $i\widetilde{S}_F(p)$ is the Feynman propagator
(\ref{sfptilde}) in momentum space related to the Dirac-like equation (\ref{tildeADE})
and $\varepsilon^\mu(\mathbf{k})$ is the photon's polarization four-vector. 
In the Lorentz covariant quantization of the photon, there are four linearly independent polarization four-vectors. 

If we use Eqs.~(\ref{U}), (\ref{uidentities}), (\ref{FPDiracTilde}), 
and (\ref{mapspinors}), we immediately get
\begin{eqnarray}
\!\!\!\mathcal{M}_a \!&\!\!=\!\!&\! -\frac{e^2}{\hbar^2}\overline{u}_D(\mathbf{p'})\slashed \varepsilon(\mathbf{k'})
iS_{D_F}\!(p+\hbar k)
\slashed \varepsilon(\mathbf{k})u_D(\mathbf{p})\!,
\label{mda}\\
\!\!\!\mathcal{M}_b \!&\!\!=\!\!&\! -\frac{e^2}{\hbar^2}\overline{u}_D(\mathbf{p'})\slashed \varepsilon(\mathbf{k})
iS_{D_F}\!(p-\hbar k')
\slashed \varepsilon(\mathbf{k'})u_D(\mathbf{p})\!.
\label{mdb}
\end{eqnarray}
Remembering that $u_D(\mathbf{p})$ 
and $iS_{D_F}(p)$ are, respectively, the standard Dirac spinor and
fermion propagator, we recognize Eqs.~(\ref{mda}) and (\ref{mdb}) as the Compton 
scattering Feynman amplitudes of standard QED \cite{man86,gre95}.

The previous example illustrates the general features present in any transition 
matrix and Feynman amplitude calculation within the theoretical framework of the 
asymmetric Dirac field QED. Point (A), with its corresponding map (\ref{mapspinors}),
will always be used to connect the asymmetric Dirac spinors with the standard ones and
to connect the Feynman propagator from both theories. Point (B) allows us to express
the conservation of energy and momentum directly in terms of the respective quantities 
for the asymmetric Dirac particles. For instance, in Eq.~(\ref{Sa2}), we could 
have written $\delta^{(4)}[(p'-\hbar \kappa)+\hbar k' - (p-\hbar \kappa) - \hbar k]$,
where $cp^0-\hbar c \kappa^0$ and $\mathbf{p}-\hbar \bm{\kappa}$ are, respectively, the energy and momentum for an asymmetric Dirac particle with wave number 
$\mathbf{p}/\hbar$. The factor $\hbar \kappa$ will always be canceled and therefore we
can simply write the conservation of energy-momentum 
directly in terms of the corresponding
quantities for the standard Dirac particles, according to the prescription given in 
point (B).

We should remark that this ``cancellation'' of the many 
$\hbar \kappa$'s in the above four-dimensional Dirac delta function, as well as
the cancellation of several exponential functions with exponents given by 
$\pm i\kappa x$ [see the discussion between Eqs.~(\ref{Sb}) and (\ref{Sa2})], are ubiquitous and not a particular feature of the Compton scattering. These 
cancellations
guarantee the validity of the crossing symmetry
principle for the asymmetric Dirac field QED. In other words,
whenever one can apply the crossing symmetry principle for the standard QED
\cite{gre03}, one can also apply it to the asymmetric Dirac field QED.

\subsubsection*{Feynman rules in momentum space}

The specific example above, as well as a general analysis similar to the ones in 
Refs.~\cite{man86,gre95}, tells us that the Feynman rules in momentum space for the 
asymmetric Dirac field QED is the same as that for standard QED, with only one exception.
Instead of the usual fermion propagator of QED in momentum space, 
we should use the propagator given by
Eq.~(\ref{sfptilde}). Being more specific, whenever we have an internal fermion line 
labeled by the four-momentum $p$, we should write the following factor, 

\begin{equation}
i\widetilde{S}_F(p)=\frac{i\hbar}{\slashed p - imc\gamma^5+i\epsilon}.
\end{equation}

And as explained above, the equivalence of any Feynman amplitude stemming from both theories is proved using the maps connecting the fermion propagator and the spinors of both theories,
Eqs.~(\ref{FPDiracTilde}) and (\ref{mapspinors}), respectively.

\subsubsection*{The cross section}

To finish the proof that both theories are empirically equivalent at the level of 
QED, we have to prove that the 
differential cross-sections derived from both theories for a given QED 
scattering process are the same. 
%
If we work through all the steps usually employed to build the differential cross
section, we identify three major logical steps \cite{man86,gre95}. The first 
two, as summarized in Ref. \cite{rig22}, are easily seen to be the same for both theories
due to points (A) and (B) given in Sec.~\ref{smatrix}. Indeed, the map of point (A)
guarantees the equivalence of the transition probability per unit time and point (B)
ensures that the conservation of energy and momentum applied to the asymmetric Dirac 
particles are properly mapped to the usual conservation of energy and momentum of 
standard Dirac particles. The last and third point \cite{man86,gre95}, as listed
in Ref. \cite{rig22}, is:
\begin{itemize}
\item The differential scattering cross section is generally defined as 
\begin{equation}
d \sigma = \frac{\mathcal{W}}{|\mathbf{j}|}\prod_{f}
\frac{Vd^3k'_{f}}{(2\pi)^3},
\label{dsigma}
\end{equation}
where $\mathcal{W}=|S_{fi}|^2/T$, 
$|\mathbf{j}|$ is the magnitude of the flux of incoming particles, and
$\prod_{f}Vd^3k'_{f}/(2\pi)^3$ represents the number of 
final scattered particles having
wave numbers between $\mathbf{k'}_{f}$ and
$\mathbf{k'}_{f} + d\mathbf{k'}_{f}$. Here $S_{fi}$ denotes 
the probability amplitude for a given process, $T$ is the duration of the experiment,
and a finite box normalization is used for the incoming and outgoing particles wave functions such that we have one particle per volume $V$. 
\end{itemize}

Equation (\ref{dsigma}) leads to the same predictions for both theories if the 
flux of incoming particles computed within each theoretical framework is the same
for a given wave number. 
The flux of particles $\mathbf{j}$ is given by Eq.~(\ref{jmu}) if we drop the charge $q$. If we use the plane wave solutions 
for the asymmetric Dirac equation as given in Sec. X of Ref.~\cite{rig23},
a direct calculation gives
\begin{equation}
\mathbf{j} = c \overline{\Psi}(x)\bm{\gamma}\Psi(x) = \frac{c^2\mathbf{p}}{V E_\mathbf{p}},
\label{j}
\end{equation}
where $\bm{\gamma}=(\gamma^1,\gamma^2,\gamma^3)$. To obtain Eq.~(\ref{j}) we normalized
$\Psi(x)$ such that $\Psi^\dagger(x)\Psi(x)=1/V$, i.e., one particle per volume $V$.

If we now compute $\mathbf{j}$ for a standard Dirac particle, using the same previous
normalization for its wave function, we also obtain Eq.~(\ref{j}). Moreover, for a standard relativistic
particle we have in SI units \cite{gre00,man86,gre95},
\begin{equation}
\mathbf{v} = \frac{c^2\mathbf{p}}{E_\mathbf{p}}.
\end{equation}
Thus, Eq.~(\ref{j}) becomes
\begin{equation}
\mathbf{v} = \mathbf{j}V,
\label{j2}
\end{equation}
which defines the velocity of a particle within a beam of particles with a 
given wave number $\mathbf{p}/\hbar$ and flux $\mathbf{j}$.

Since we have just seen that in both theories the flux $\mathbf{j}$ is the same,
Eq.~(\ref{j2}) implies that the asymmetric and standard Dirac particles have the 
same velocity $\mathbf{v}$. But $\mathbf{v}=\mathbf{p}/m$ for a standard relativistic
particle, which implies that $\mathbf{p}/m$ is also the velocity 
for an asymmetric Dirac particle sharing the same wave number with the corresponding 
standard Dirac particle. This is the justification of point (C) given in Sec.~\ref{smatrix}
and the proof that the 
differential cross section (\ref{dsigma}) are equivalent for both theories.

Note that the above differential cross section is the polarized one, where we have
not summed or averaged over the fermion spins or photon polarizations. However,
since Eq.~(\ref{dsigma}) are equivalent in both theories, we can always first convert it
to its standard QED analog and only then sum and average 
over all fermion spins and photons polarizations. This proves the equivalence of the 
unpolarized cross sections as well.

If we choose, nevertheless, to do all calculations directly using the asymmetric Dirac spinors, the computation of the unpolarized cross sections are slightly different since the  ``spin sums rules'' \cite{man86,gre95} have to be adapted to the appropriate context. Of
course, the final results will be the same, as the above discussion shows. 
We should also note that the ``polarization sum rules'' for the photons are easily seen to be the same of standard QED. This is true since 
the electromagnetic part of the asymmetric Dirac field QED are exactly the same of the 
standard one; we have adopted the same Lorentz covariant second quantization recipe 
for the photon field in both theories.

In order to show how to deal with the fermion spin sums directly using the asymmetric Dirac spinors, and without losing in generality, let us fix our attention on Compton scattering. The unpolarized cross-section for Compton scattering is proportional to \cite{man86} 
\begin{equation}
X = \frac{1}{2}\sum_{r=1}^{2}\sum_{s=1}^{2}|\mathcal{M}|^2,
\label{x}
\end{equation}
where $\mathcal{M}$ is the Feynman amplitude for this process. The Feynman amplitude
can be generically written as \cite{man86}
\begin{equation}
\mathcal{M} = \overline{u}_s(\mathbf{p'})\Gamma u_r(\mathbf{p}),
\label{mx}
\end{equation}
where $\Gamma$ is an arbitrary $4\times 4$ matrix made of products of the usual 
gamma matrices.

Following \cite{man86}, we can write Eq.~(\ref{x}) as follows,
\begin{equation}
X \!=\! \frac{1}{2}\!\left[\!\sum_{s=1}^{2}u_{s_\delta}(\mathbf{p'})
\overline{u}_{s_\alpha}(\mathbf{p'})\!\right]\!
\Gamma_{\alpha\beta}\! 
\left[\!\sum_{r=1}^{2}u_{r_\beta}(\mathbf{p})
\overline{u}_{r_\gamma}(\mathbf{p})\!\right]\!
\widetilde{\Gamma}_{\gamma\delta},
\label{x2}
\end{equation}
where 
\begin{equation}
\widetilde{\Gamma} = \gamma^0\Gamma^\dagger \gamma^0.
\end{equation}
Note that in Eq.~(\ref{x2}) we also sum over repeated spinor indexes. 

So far, what is shown in Eq.~(\ref{x2}) is formally the same to what we get when 
dealing with standard Dirac spinors. The difference shows up when we use 
Eq.~(\ref{uubar}), or Eq.~(\ref{vvbar}) when we have $v_r(\mathbf{p})$
instead of $u_r(\mathbf{p})$. Employing Eqs.~(\ref{uubar}) and (\ref{lpm}) we get
\begin{equation}
X = \frac{1}{2}\mbox{Tr} \left[\frac{\slashed p' - imc \gamma^5}{2mc}\Gamma 
\frac{ \slashed p - imc \gamma^5}{2mc}\widetilde{\Gamma} \right].
\label{x3}
\end{equation}
If we now use Eq.~(\ref{uidentities}) and the cyclic property of the trace, we can write Eq.~(\ref{x3}) as follows,
\begin{equation}
X = \frac{1}{2}\mbox{Tr} \left[\frac{\slashed p' + mc}{2mc}U\Gamma U 
\frac{ \slashed p + mc}{2mc}U\widetilde{\Gamma}U \right].
\label{x4}
\end{equation}

Comparing Eq.~(\ref{x4}) with the corresponding one for the standard QED \cite{man86},
we realize that the former can be obtained from the latter by the following substitutions,
\begin{eqnarray}
\Gamma \rightarrow U\Gamma U, & \widetilde{\Gamma} \rightarrow U\widetilde{\Gamma}U. 
\end{eqnarray}
Equation (\ref{x4}) together with three other similar ones, obtained by all different
arrangements of the spinors $u$ and $v$ in Eq.~(\ref{mx}), illustrate the general 
way of handling spin sums directly within the theoretical framework of the asymmetric
Dirac field QED.

Incidentally, if we observe the computation of the unpolarized cross sections for the 
several QED processes as given in Ref.~\cite{man86}, namely, lepton pair creation in electron-positron collisions, Bhabha scattering, Compton scattering where either we 
sum and average
over the electron spins only or we sum and average over both 
the electron spins and photon
polarizations, and the scattering of a fermion by an external electromagnetic field, 
we observe that 
$\Gamma$ is either $\gamma^\mu$ or $\gamma^\mu\gamma^\alpha\gamma^\nu$. 

However, whenever $\Gamma$ is given by the product of an odd number of
the gamma matrices $\gamma^0, \gamma^1, \gamma^2,$ and $\gamma^3$ as, for instance,
in the scattering processes listed above, a direct calculation gives
\begin{eqnarray}
U\Gamma^{\mbox{\tiny{odd}}} U =\Gamma^{\mbox{\tiny{odd}}}, & 
U\widetilde{\Gamma}^{\mbox{\tiny{odd}}} U =\widetilde{\Gamma}^{\mbox{\tiny{odd}}}.
\label{oddgamma}
\end{eqnarray}
Equation (\ref{oddgamma}) when inserted into Eq.~(\ref{x4}) leads exactly to the 
corresponding expression for the standard QED. 
This is yet another way of seeing the equivalence of 
the unpolarized cross sections of the asymmetric Dirac field QED and the standard QED.

Moreover, since $U$ commutes with $\gamma^5$, we also have
\begin{eqnarray}
U\gamma^5\Gamma^{\mbox{\tiny{odd}}} U =\gamma^5\Gamma^{\mbox{\tiny{odd}}}, & 
U\gamma^5\widetilde{\Gamma}^{\mbox{\tiny{odd}}} U 
= \gamma^5\widetilde{\Gamma}^{\mbox{\tiny{odd}}},
\label{oddgamma5}
\end{eqnarray}
which helps in proving the equivalence of Eq.~(\ref{x4}) for both theories when we 
compute cross sections for electrons with particular helicity states \cite{man86}.


We should also mention that the equivalence between the standard QED and the asymmetric
Dirac field QED goes beyond the tree level. Using Eqs.~(\ref{FPDiracTilde}) and (\ref{mapspinors}), and employing the prescriptions (A)-(C) of Sec.~\ref{smatrix}, 
we can map any Feynman amplitude and cross section from one theory to the corresponding
ones of the other theory describing the same physical process. As such,
higher order terms in perturbation theory describing a given scattering process, as
well as the computation of the radiative corrections and the renormalization techniques
from standard QED, can all be readily carried over to the asymmetric Dirac field QED,
leading to the same results and predictions.

\section{Beyond quantum electrodynamics}
\label{bqed}

The arguments used in Sec.~\ref{smatrix} to prove the equivalence between the standard and
the asymmetric Dirac field QED can be straightforwardly carried over to analyze 
the possible equivalence of both theories when we employ a different type of interaction.
Since the standard rules to compute the cross section for both theories are essentially
the same if we adopt the prescriptions (A), (B), and (C) given in Sec.~\ref{smatrix}, the 
proof of the equivalence can be traced back to proving whether or not the interaction 
Hamiltonian density are equivalent in both theories.

As shown in Sec.~\ref{qed}, the interaction Hamiltonian density of QED, given by Eq.~(\ref{hinta}), is the same in both theories. This comes about because of the
particular transformation rules connecting the fields of both theories, Eqs.~(\ref{aDtoD}) and (\ref{aDtoDadj}), and because of 
the way $\gamma^\mu$ responds to those transformation rules.
Specifically, the first mathematical identity given by Eq.~(\ref{uidentities})
is crucial to prove the equality of the interaction Hamiltonian in both theories.
In other words, Eq.~(\ref{hinta}) is invariant under the action of Eqs.~(\ref{aDtoD})
and (\ref{aDtoDadj}) because of Eq.~(\ref{uidentities}).

Our first goal in this section is to investigate how the
other Dirac bilinears respond to those transformation rules. As we will see, not all 
bilinears are invariant under the action of Eqs.~(\ref{aDtoD})
and (\ref{aDtoDadj}). Our second goal is to investigate what kinds of interactions 
involving scalar and spinor fields, or different spinor fields, 
are allowed when we demand that the Lorentz symmetry be fully respected. 
We will see that the requirement for Lorentz invariant interactions restricts
severely the types of interactions allowed for the asymmetric Dirac fields, a 
constraint that is stronger than the ones required for the standard Dirac fields.
Finally, we will end this section investigating, in a very qualitative way, what 
might happen when we include interaction terms coming from massive and 
possibly non-abelian Gauge fields that transform under a proper Lorentz transformation
similarly to the asymmetric Dirac fields. 
Again, the demand that we must have Lorentz invariant 
Hamiltonian densities leads to stringent constraints on the possible types of allowed interactions.

\subsection{Dirac bilinears}

If we use Eqs.~(\ref{aDtoD}), (\ref{aDtoDadj}), and (\ref{uidentities}), we readily obtain
\begin{eqnarray}
\overline{\Psi}\Psi &\longrightarrow& -\overline{\Psi}_Di\gamma^5\Psi_D, \label{b1}\\
\overline{\Psi}i\gamma^5\Psi &\longrightarrow& \overline{\Psi}_D\Psi_D, \label{b2}\\
\overline{\Psi}\gamma^\mu\Psi &\longrightarrow& \overline{\Psi}_D\gamma^\mu\Psi_D,
\label{b3} \\
\overline{\Psi}\gamma^5\gamma^\mu\Psi &\longrightarrow& 
\overline{\Psi}_D\gamma^5\gamma^\mu\Psi_D,
\label{b4} \\
\overline{\Psi}\sigma^{\mu\nu}\Psi &\longrightarrow& 
-\overline{\Psi}_Di\gamma^5\sigma^{\mu\nu}\Psi_D \nonumber \\ 
&&= -\frac{1}{2}\epsilon^{\mu\nu\alpha\beta}
g_{\eta\alpha}g_{\lambda\beta}\overline{\Psi}_D\sigma^{\eta\lambda}\Psi_D.
\label{b5}
\end{eqnarray}
Here $\epsilon^{\alpha'\beta'\mu'\nu'}=
g^{\alpha'\alpha}g^{\beta'\beta}g^{\mu'\mu}g^{\nu'\nu}\epsilon_{\alpha\beta\mu\nu}$
and $\epsilon_{\alpha\beta\mu\nu}$ is the completely antisymmetric four-dimensional 
Levi-Civita symbol, i.e., $\epsilon_{\alpha\beta\mu\nu}=1$ for 
$(\alpha,\beta,\mu,\nu)=(0,1,2,3)$ and even permutations while 
$\epsilon_{\alpha\beta\mu\nu}=-1$ for odd permutations.

Looking at Eqs.~(\ref{b1})-(\ref{b5}), we realize that only the bilinears given by
Eqs.~(\ref{b3}) and (\ref{b4}) have the same form when we go from the asymmetric Dirac
fields to the standard ones. Equation (\ref{b3}) is the bilinear associated with QED,
namely, the term proportional to the electric current that couples with the 
electromagnetic abelian gauge field describing the photon.  
The remaining three terms, Eqs.~(\ref{b1}), (\ref{b2}),
and (\ref{b5}), are mapped to different bilinears though. 

For instance, an interaction term proportional to $(\overline{\Psi}\Psi)^n$, with 
$n>0$ being an integer, will not lead to the same predictions if we work with the 
standard Dirac fields and an interaction term proportional to 
$(\overline{\Psi}_D\Psi_D)^n$. Actually, according to Eq.~(\ref{b1}),
the predictions stemming from $(\overline{\Psi}\Psi)^n$ in the context of the 
asymmetric Dirac fields 
will be the same as those coming from $(-1)^n(\overline{\Psi}_Di\gamma^5\Psi_D)^n$
in the context of the standard Dirac fields. Depending on the value of $n$, an 
attractive interaction in the  context of the asymmetric Dirac fields can be mapped to
a repulsive one when we move to the standard Dirac fields. 
For example, when $n=2$, $(\overline{\Psi}\Psi)^2$ is mapped to 
$-(\overline{\Psi}_D\gamma^5\Psi_D)^2$ while when $n=4$ we have
$(\overline{\Psi}\Psi)^4$ going to $(\overline{\Psi}_D\gamma^5\Psi_D)^4$. The minus
sign in the former case indicates a change in the sign of the coupling constant, a feature
that is not seen in the latter case.

\subsection{More than one spinor field}
\label{twospniors}

We now assume, for definiteness, that we are dealing with two kinds of leptons. 
An electron and a different lepton, for instance, a muon or a tau. The asymmetric electron spinor will be denoted by $\Psi_e$ while the other lepton spinor by $\Psi_l$. The 
corresponding standard Dirac spinors will be denoted by  $\Psi_{D_e}$ and 
$\Psi_{D_l}$, respectively. 

Since different leptons have different masses, the asymmetric
spinors and standard Dirac spinors are connected by 
[cf. Eqs.~(\ref{aDtoD}) and (\ref{aDtoDadj})],
\begin{eqnarray}
\Psi_j(x) &=& e^{i\kappa_j x}U\Psi_{D_j}(x),
\label{aDtoDL} \\
\overline{\Psi}_j(x) &=& e^{-i\kappa_j x} \overline{\Psi}_{D_j}(x)U,
\label{aDtoDadjL}
\end{eqnarray}
where $j$ stand for $e$, an electron, or $l$, the other lepton. Note that in the 
previous equations, $\kappa_j x =(\kappa_j)_\mu x^\mu 
=(\kappa_j)_0x^0+(\kappa_j)_1x^1+(\kappa_j)_2x^2
+(\kappa_j)_3x^3$ and $(\kappa_j)^2=(\kappa_j)_\mu(\kappa_j)^\mu=m_jc/\hbar$,
with $m_j$ being either the mass of the electron or the mass of the other lepton
[cf. Eqs.~(\ref{restmass}) and (\ref{dispKG})].

For standard Dirac spinors, an interaction term proportional to 
\begin{equation}
(\overline{\Psi}_{D_l}\Psi_{D_e})^n \label{psibarpsin}
\end{equation}
is perfectly legitimate. Here and in what follows the corresponding Hermitian conjugate term will be omitted for simplicity. And, as always, normal ordering is implicitly assumed. 
This interaction term is a Lorentz scalar under a 
proper Lorentz transformations since both $\Psi_{D_e}$ and $\Psi_{D_l}$ transform 
in exactly the same way after a proper Lorentz transformation \cite{gre00,man86,gre95}.
Using the notation introduced in Sec.~\ref{asd}, this means that
\begin{eqnarray}
\Psi'_{D_j}(x') &=& S \Psi_{D_j}(x), \label{psij}\\
\overline{\Psi}'_{D_j}(x') &=&\overline{\Psi}_{D_j}(x)S^{-1}, \label{psibarj}
\end{eqnarray}
where $j=e,l$. Equations (\ref{psij}) and (\ref{psibarj}) when inserted into 
Eq.~(\ref{psibarpsin}) clearly show that we have an interaction term respecting the 
Lorentz symmetry (invariant under a proper Lorentz transformation).

On the other hand, for the asymmetric Dirac spinors,
an interaction proportional to 
\begin{equation}
(\overline{\Psi}_{l}\Psi_{e})^n \label{psibarpsinA}
\end{equation}
is not a Lorentz scalar under a 
proper Lorentz transformation since now $\Psi_{e}$ and $\Psi_{l}$ transform 
differently after it. This happens because the 
transformation rule for $\Psi_{j}$ depends on the value of $(\kappa_j)^\mu$,
as we realize looking at Eqs.~(\ref{Minf})-(\ref{Mfinite}), and each lepton has
a different set of values for $(\kappa_j)^\mu$.

By making use of Eqs.~(\ref{aDtoDL}) and (\ref{aDtoDadjL}), we can also understand why an interaction modeled by Eq.~(\ref{psibarpsinA}) violates the Lorentz symmetry. 
Inserting Eqs.~(\ref{aDtoDL}) and (\ref{aDtoDadjL}) into
Eq.~(\ref{psibarpsinA}) we get
\begin{equation}
(\overline{\Psi}_{l}\Psi_{e})^n = e^{in(\kappa_e - \kappa_l)x}
(-1)^n(\overline{\Psi}_{D_l}i\gamma^5\Psi_{D_e})^n.
\label{psibarpsinD}
\end{equation}
The term $(-1)^n(\overline{\Psi}_{D_l}i\gamma^5\Psi_{D_e})^n$ at the right hand side of Eq.~(\ref{psibarpsinD})
is clearly an invariant under proper Lorentz transformations but 
$e^{in(\kappa_e - \kappa_l)x}$ is not since $(k_j)^\mu$ is not a four vector. 
This clearly shows that the Lorentz symmetry is broken for that type of interaction.
Note that interactions given by $(\overline{\Psi}_{j}\Psi_{j})^n$, where only spinors
of the same flavor are present, respect the Lorentz symmetry. In this case
$\kappa_e - \kappa_l \rightarrow \kappa_j - \kappa_j=0$, which means that 
the Lorentz symmetry offending term is no longer present.

Therefore, we can only have an interaction involving different flavors and that respects the Lorentz 
symmetry if it is given by
\begin{equation}
(\overline{\Psi}_{e}\Psi_{e})^{n_1}(\overline{\Psi}_{l}\Psi_{l})^{n_2}, 
\label{ppbar}
\end{equation}
where $n_1$ and $n_2$ are arbitrary positive integers. 

The above analysis used the bilinear (\ref{b1}) as a concrete example. However, it 
is also valid for the other four types of bilinears. Similar arguments can be used
to restrict the types of interactions built from the other four bilinears, 
Eqs.~(\ref{b2})-(\ref{b5}), or combinations therefrom.

\subsection{Complex scalar or gauge fields}

Let us now investigate what types of interactions are allowed when, in addition to the
two spinor fields above, we also have one complex scalar field. If we denote by 
$\Phi_{KG}$ a standard complex Klein-Gordon field, it is obvious that the following
interaction term is Lorentz invariant,
\begin{equation}
(\Phi_{KG})^{n_s}(\overline{\Psi}_{D_l}\Psi_{D_e})^n, \label{phipsi}
\end{equation}
where $n_s$ and $n$ are arbitrary positive integers. The analogous term involving the 
asymmetric fields, namely, the asymmetric Dirac spinors \cite{rig23} and the Lorentz covariant Schr\"odinger fields \cite{rig22}, is not in general a Lorentz invariant. 
The reason for this is related to the fact that these asymmetric fields have 
transformation rules under a proper Lorentz transformation that depend on the 
four invariants given by $(\kappa_j)^\mu$. This dependence shows up as a phase that is a function of the space-time variables and of $(\kappa_j)^\mu$. 
And since the invariants $(\kappa_j)^\mu$ depend on the values of the rest masses associated with the interacting particles,
in general the phases will not cancel out each other when we deal with particles having different masses.

Similarly to what we did in Sec.~\ref{twospniors}, we can understand the above point if
we use Eqs.~(\ref{aDtoDL}), (\ref{aDtoDadjL}), and the corresponding one connecting 
the Lorentz covariant Schr\"odinger fields with the standard complex Klein-Gordon ones \cite{rig22}, 
\begin{equation}
\Phi(x) = e^{i\kappa_s x}\Phi_{KG}(x).
\label{GLStoKG}
\end{equation}
Here $(\kappa_s)^2=(\kappa_s)_\mu(\kappa_s)^\mu=m_sc/\hbar$,
with $m_s$ being the mass of the scalar particle described by $\Phi(x)$.

If we use Eqs.~(\ref{aDtoDL}), (\ref{aDtoDadjL}), and (\ref{GLStoKG}), the following 
interaction term in the context of the asymmetric fields can be written as 
\begin{eqnarray}
\hspace{-1cm}(\Phi)^{n_s}(\overline{\Psi}_l\Psi_e)^n &=& 
e^{i[n(\kappa_e - \kappa_l)+n_s\kappa_s]x} \nonumber \\
&&\times (-1)^n(\Phi_{KG})^{n_s}(\overline{\Psi}_{D_l}i\gamma^5\Psi_{D_e})^n. \label{phipsiA}
\end{eqnarray}
The expression in the last line of Eq.~(\ref{phipsiA}) is clearly invariant under a proper 
Lorentz transformation. Thus, we can only make the whole expression invariant if
\begin{equation}
n(\kappa_e - \kappa_l)+n_s\kappa_s = 0. \label{constraint1}
\end{equation}

Equation (\ref{constraint1}) tells us that only certain combinations of the positive integers $n$ and $n_s$ may lead to a Lorentz invariant interaction. Indeed, each 
set of four invariants $(\kappa_j)^\mu$, with $j=e,l,s$, is determined in such a way
that it is compatible with the rest mass $m_j$ of the $j$-th particle involved in the interaction above.
With the values of $(\kappa_j)^\mu$ fixed in this way, 
we have to choose $n$ and $n_s$ such that 
Eq.~(\ref{constraint1}) is satisfied, guaranteeing an interaction that fully respects the 
Lorentz symmetry.

The argument above can 
be extended, or better yet, reversed, if instead of a scalar particle
we have a massive complex gauge field (abelian or non-abelian) mediating the interaction 
between the two spinors. First, if it were possible to consistently develop a complex gauge field theory whose fields transform under a proper Lorentz
transformation similarly to the asymmetric spinor and complex scalar fields studied here.
Second, if we could consistently apply a ``minimal coupling prescription'' leading to the 
interaction between the asymmetric spinor and complex gauge fields, then
a similar constraint given by Eq.~(\ref{constraint1}), that guarantees the Lorentz invariance of the interaction (\ref{phipsiA}), could be used to determine the mass of the gauge boson. 

This would be the case because the interaction term would be fixed by the minimal 
coupling prescription. We no longer would be free to choose the interaction term.
In other words, the values of $n_s$ and $n$ in the 
corresponding interaction term would not be arbitrary but determined by the specific form of the interaction arising after the minimal coupling prescription is applied. 
As such, we would have
to choose the values of $(\kappa_j)^\mu$ in order to satisfy a similar constraint as
given by Eq.~(\ref{constraint1}). And since the values of $(\kappa_j)^\mu$ determine
the mass of particle $j$, the mass of the gauge boson would be fixed once the masses 
of the spinors enter as input. Note that theories built along these lines may also provide 
a fresh outlook in the understanding of the flavor puzzle \cite{fer15}.

Note that taking the previous argument at face value might lead to a electroweak theory 
that does not respect the lepton flavor universality \cite{gre00b}. Indeed, to achieve universality all leptons should have the same coupling constant with a given
boson and the interaction term should have the same functional form. 
Since we have three generations of leptons, we will run out of parameters to satisfy three similar expressions as given by Eq.~(\ref{constraint1}). However, the electroweak theory is
non-abelian and it is expected that for non-abelian gauge fields the phase of 
Eq.~(\ref{GLStoKG}) will be generalized to a space-time dependent unitary matrix, 
whose dimension is related to the internal space dimension 
from which the non-abelian theory is built. This will eventually lead to more parameters than the ones appearing in the exponential of Eq.~(\ref{phipsiA}). As such, it is not unlikely that we may have enough parameters to determine the boson masses and, at the same time, keep lepton flavor universality. The definitive answer to this question can only be given
when (and if) a non-abelian asymmetric field theory is fully developed.

\section{Conclusion}
\label{con}

We showed that it is possible to consistently second quantize 
the asymmetric Dirac fields, which were introduced at the first quantization level 
in Ref.~\cite{rig23}. By applying the electromagnetic minimal coupling
prescription to the free field Lagrangian density describing the asymmetric Dirac fields,
a fully renormalizable QED-like quantum field theory consistently emerged. Moreover,
by properly choosing the parameters characterizing the free asymmetric spinorial 
fields, we proved that the obtained QED-like theory gives exactly the same
predictions of the
standard QED. This latter result is far from trivial and could not have been anticipated
by simply inspecting the assumptions employed in the construction of the 
asymmetric scalar and Dirac fields \cite{rig22,rig23}. 

Indeed, we can list the following four main features that set apart the present theory,
and the corresponding scalar one given in Ref.~\cite{rig22}, from the standard ones. 
First, the Lagrangian densities
and the associated scalar and spinorial wave equations describing those asymmetric
fields are formally different from the corresponding quantities related to the standard Klein-Gordon and Dirac fields. Second, the transformation rules for the asymmetric fields under a proper Lorentz transformation are fundamentally different from the rules for 
standard scalar and spinor fields. The asymmetric fields transformation rules, contrary
to the standard ones, have a space-time dependent phase in addition to the usual rules.
Third, the masses of the asymmetric particles are functions of four relativistic invariants
that naturally appear in the free field Lagrangian densities. And these four 
relativistic invariants change sign after improper Lorentz transformations or after 
the charge conjugation operation in the same way as the components of a four-vector.
The mass $m$, though, is invariant under those same transformations since it depends 
quadratically on those invariants. Fourth, particles and antiparticles no longer have
degenerate energies and momenta when sharing the same wave number. In its simplest
formulation, we can build an asymmetric quantum field theory such that particles and antiparticles have the same momenta and a 
gap of $2mc^2$ in their energies while sharing the same wave number. Despite 
all these unusual features, the asymmetric classical and quantum field theories 
developed here and in Refs.~\cite{rig22,rig23} respect the Lorentz symmetry and
can be made equivalent to the standard Klein-Gordon and Dirac theories when 
minimally coupled to electromagnetic fields.

The next natural question is the extension of the present ideas, and those of 
Refs.~\cite{rig22,rig23}, to massive gauge fields that transform under a proper 
Lorentz transformation similarly to the way the asymmetric scalar and Dirac fields
do \cite{rig22,rig23}. In other words, can we build a 
renormalizable electroweak theory whose gauge fields but one (the photon field) 
are massive from the start, fully respects the Lorentz symmetry, and reproduces the predictions of the standard
electroweak theory? Moreover, can we explore the asymmetry of the fields, namely, the 
gap in the energy between particles and antiparticles in order to provide a viable 
explanation for the asymmetry between matter and antimatter in our universe in a 
fully relativistic theory \cite{din04}? Since antiparticles are 
more energetic than particles for a given wave number, it is not unlikely that a 
electroweak-like theory built on these principles may lead to antiparticles being more
unstable than particles. And the CP violation, would it be stronger for those 
asymmetric fields? Would we get new sources of CP violation in this theoretical 
framework \cite{bra12,nis06,nis12,nis13,nis15}? 
What about the strong force, can a theory describing it be formulated along the 
principles outlined here?

Another interesting question that deserves further exploration is related to the
gravitational force. As qualitatively discussed here and 
in Refs.~\cite{rig22,rig23}, the fact that 
the inertial mass of a particle, in the theoretical framework of the asymmetric fields,
is a quadratic function of four relativistic invariants, 
suggests that the coupling of matter 
with the gravitational field can be implemented via those invariants instead of the 
mass itself. 
This opens the possibility to model the gravitational interaction between particles
and antiparticles differently. It is possible to build a theory in which 
a particle and an antiparticle repel each other gravitationally while still keeping the masses and energies of particles and antiparticles positive \cite{rig22,rig23}. In this theory
particles attract particles, antiparticles attract antiparticles, with the repulsive
force only between a particle and an antiparticle \cite{rig22,rig23}.
Would this approach to model
the gravitational interaction lead to a consistent renormalizable quantum field theory of
gravity? Would its predictions be consistent with the experimental data we
have collected so far concerning the interaction of matter and antimatter 
with a gravitational field?

Finally, the ideas presented here and in Refs. \cite{rig22,rig23}
show that it is possible to consistently build relativistic classical 
and quantum field theories whose scalar and spinorial fields transform in 
a non-standard way under proper Lorentz 
transformations. In spite of that, these theories can be 
made to reproduce exactly the experimental predictions of standard
Klein-Gordon and Dirac fields interacting with photon fields,
with the electromagnetic interaction being introduced
via the minimal coupling prescription.  Two other peculiar aspects of the 
asymmetric field theories developed here and Refs. \cite{rig22,rig23} lie in the 
non-degeneracy of the energy and momentum for a particle and antiparticle sharing 
the same wave number and the more stringent constraints on the possible types of 
interactions among those fields that fully respect the Lorentz symmetry. 
The non-degeneracy in the energy for particles and antiparticles does not lead to experimental predictions beyond the standard model at the level of QED, as we proved
here. However, it is not unlikely that the formulation of a 
fundamental theory where flavor is not conserved (electroweak theory) in terms
of asymmetric spinorial and gauge fields 
may lead to an experimental prediction beyond the standard model.
In electroweak-like theories, the decay of particles is a fundamental aspect of those
theories and the non-degeneracy in the energy of particles and antiparticles may
manifest itself in different decay rates for particles and antiparticles sharing the
same mass.

\acknowledgments
The author thanks the Brazilian agency CNPq
(National Council for Scientific and Technological Development) for partially
funding this research.

\end{document}